\documentclass[11pt,aps,a4paper,eqsecnum,amsmath,amssymb,
  notitlepage,nofootinbib]{revtex4-1}

\usepackage{graphicx}
\usepackage{bbm}
\usepackage{booktabs}

 \newcommand{\klammer}[1]{\left( #1\right)}
 \newcommand{\gklammer}[1]{\left\{ #1\right\}}
 \newcommand{\eckklammer}[1]{\left[ #1\right]}
 
 \newcommand{\ket}[1]{|#1\rangle}

 \newcommand{\End}{\operatorname{End}}
 \newcommand{\parity}{\operatorname{p}}
 \newcommand{\sTP}{\otimes_{\hspace{-.1em}s}}

\newcommand{\opc}[1]{c^{\phantom{\dagger}}_{#1}}
\newcommand{\opcd}[1]{c^{\dagger}_{#1}}
\newcommand{\opn}[1]{n^{\phantom{\dagger}}_{#1}}
\newcommand{\opnq}[1]{\bar{n}^{\phantom{\dagger}}_{#1}}
 \newcommand{\eins}{{\mathbbm{1}}}
 \newcommand{\calA}{\mathcal{A}}
 \newcommand{\calB}{\mathcal{B}}
 \newcommand{\calC}{\mathcal{C}}
 \newcommand{\calD}{\mathcal{D}}
 \newcommand{\calT}{\mathcal{T}}

 \newcommand{\str}[1]{\mathrm{str}\left\{\, #1 \,\right\}}
 \newcommand{\sTr}[2]{ \mathrm{str}_{_{#1}}\left\{ \, #2\,\right\}}

 \newcommand{\st}{\mathrm{st}}
 \newcommand{\ist}{\mathrm{ist}}
 \newcommand{\sT}[1]{\mathrm{st}_#1}
 \newcommand{\isT}[1]{\mathrm{ist}_#1}

 \newcommand{\sss}[1]{{\scriptscriptstyle #1}}
 \newcommand{\herv}[1]{\textit{#1}}
 \newcommand{\sine}[3][]{\operatorname{s}^{#1}_{#2}(#3)}
 \newcommand{\comment}[1]{}

\begin{document}
\preprint{}
\title{Truncation identities for the small polaron fusion hierarchy} 

\author{Andr\'e M. Grabinski}

\author{Holger Frahm}

\affiliation{%
Institut f\"ur Theoretische Physik, Leibniz Universit\"at Hannover,
Appelstra\ss{}e 2, 30167 Hannover, Germany}


\begin{abstract}
  We study a one-dimensional lattice model of interacting spinless fermions.
  This model is integrable for both periodic and open boundary conditions, the
  latter case includes the presence of Grassmann valued non-diagonal boundary
  fields breaking the bulk $U(1)$ symmetry of the model.  Starting from the
  embedding of this model into a graded Yang-Baxter algebra an infinite
  hierarchy of comuting transfer matrices is constructed by means of a fusion
  procedure.  For certain values of the coupling constant related to
  anisotropies of the underlying vertex model taken at roots of unity this
  hierarchy is shown to truncate giving a finite set of functional equations
  for the spectrum of the transfer matrices.  For generic coupling constants
  the spectral problem is formulated in terms of a TQ-equation which can be
  solved by Bethe ansatz methods for periodic and diagonal open boundary
  conditions.  Possible approaches for the solution of the model with generic
  non-diagonal boundary fields are discussed.
\end{abstract}

\maketitle

%
%
%
\section{Introduction}
The small polaron model provides an effective description of the behaviour of
an additional electron in a polar crystal \cite{FeYa78,MaFe84}.  In one
spatial dimension this lattice system of interacting spinless fermions can be
constructed within the framework of the Quantum Inverse Scattering Method
\cite{QISM} allowing to compute the excitation spectrum by Bethe ansatz
techniques, see e.g.\ \cite{PuZh86,ZhJW89}.  By means of a graded
generalization \cite{MeNe91,Gonz94,BGZZ98} of Sklyanin's reflection algebra
\cite{Skly88} it was possible to provide the small polaron model with open
boundary conditions while keeping
its integrability intact.  These integrable boundary conditions are encoded in
$c$-number valued $2\times2$-matrix solutions to the reflection equations
\cite{Zhou96a,Zhou97,GuGR98}.

Diagonal boundary matrices correspond to boundary chemical potentials in the
Hamiltonian.  In this case the small polaron model is equivalent to the
spin-1/2 XXZ Heisenberg chain with boundary magnetic fields by means of a
Jordan-Wigner transformation, similarly as in the case of periodic boundary
conditions where this equivalence holds up to a boundary twist depending on
the particle number \cite{PuZh86,UmFW99}.  As a consequence the spectrum of
the open small polaron model can be obtained using Bethe ansatz methods
\cite{UmFW99,GuFY99,WaFG00}.
For general non-diagonal solutions to the reflection equations this
equivalence does not hold as a consequence of the non-local nature of the
Jordan-Wigner transformation.  Furthermore, the underlying grading implies
that solutions to the reflection equations for the small polaron model are
\emph{super matrices} \cite{CORNWELL3}.  In the corresponding Hamiltonian the
resulting additional boundary terms do not conserve particle number and have
anti-commuting scalars, i.e.\ odd Grassmann numbers, as amplitudes.  The fact
that the $U(1)$ symmetry of the model is broken implies that in general there
is no simple eigenstate (e.g.\ the Fock vacuum) of the model which can be used
as a reference state for the algebraic Bethe ansatz.  Therefore, alternative
approaches such as functional Bethe ansatz methods have to be employed to
analyze the spectrum of the model.
This situation is, in fact, very similar to the case of non-diagonal boundary
magnetic fields in the spin-$1/2$ Heisenberg chains: in the approaches used so
far the solution of the spectral problem relies on constraints between the
boundary fields at the two ends of the chain or restrictions on the
anisotropy, or it is limited to small finite systems thereby reducing their
usefulness to study this system in the thermodynamic limit
\cite{Nepo02,CaoX03,Nepo04,YaNZ06,BaKo07,Galleas08,FrSW08,FGSW11,Nicc12}.

In a previous publication \cite{GrFr10} we have investigated the applicability
of Bethe ansatz methods in the simpler case of a model of free fermions with
similar open boundary conditions.  We found that for a certain class of
non-diagonal boundary super matrices, a unitary transformation on the
auxiliary space allowed for an exact solution of the free fermion model.
Furthermore, the functional equations obtained there could be easily
generalized to describe the spectrum of the model for arbitrary non-diagonal
boundary fields.  Unfortunatley, this approach can not be applied directly to
the small polaron model.

In the present paper we initiate a study as to whether the nilpotency of the
off-diagonal boundary parameters in a graded model allows to bypass some of
the problems arising in the case of the spin-1/2 XXZ chain with non-diagonal
boundary fields.
Following ideas \cite{Nepo02,YaNZ06} developed in the context of the spin-1/2
XXZ Heisenberg chain and later generalized to the XYZ chain \cite{YaZh06} and
integrable higher spin XXZ models \cite{FrNR07} we adapt the fusion procedure
\cite{KuRS81,MeNe92a,Zhou96b} for the transfer matrix of the quantum chain to
the graded case of the small polaron model.
We derive the fusion hierarchy of functional equations for a commuting family
of transfer matrices for the small polaron model.  Assuming the existence of a
certain limit we formulate the spectral problem of this model for periodic
and general open boundary conditions in terms of functional TQ-equations.  For
periodic and diagonal open boundary conditions these equations are shown to
coincide with the known result obtained from using the algebraic Bethe ansatz.
For special values of the interaction parameter related to roots of unity of
the anisotropy parameter we derive truncation identities for the fusion of the
relevant objects, in particular the transfer matrices.  Using these identities
the fusion hierarchy reduces to a set of relations between finitely many
quantities.

%
\section{The small polaron as a fundamental integrable model}
\label{ch:SmallPolaronModel}
Some materials exhibit a strong electron-phonon coupling that considerably
reduces the mobility of electrons within the conduction band. This interaction
may be regarded as an increase of the electron's effective mass, thus giving
raise to quasi-particles called \herv{polarons}. If the electron is
essentially trapped at a single lattice-site the corresponding quasi-particle
is said to be a \herv{small polaron}. In this case, electron transport occurs
either by thermally activated hopping (at high temperatures) or by tunneling
(at low temperatures). 

In the case of periodic boundary conditions (PBC) the $N$-site small polaron
model is characterized by the Hamiltonian 
\begin{equation}
 H^\sss{\text{PBC}} = \sum_{j=1}^{N} H_{j,j+1}\qquad \text{with}\qquad
 H_{N,N+1} \equiv H_{N, 1} 
 \label{eq:PBCHamiltonian}
\end{equation}
with a Hamiltonian density $H_{j,j+1}$ defined as
\begin{equation}
 H_{j,j+1}=-t\klammer{\opcd{j+1}\opc{j}+\opcd{j}\opc{j+1}}+
 V\klammer{\opn{j+1}\opn{j}+\opnq{j+1}\opnq{j}}  
 \label{eq:HamiltonianDensity}
\end{equation}
where $\opcd{k}$ and $\opc{k}$ label the creation resp. annihilation operators
of spinless fermions at site $k$, which are subject to the anticommutation
relations $[\opcd{\ell},\opc{k}]_\sss{+} = \delta_{\ell k}$. Moreover, it is
convenient to define number operators 
$\opn{k} \equiv \opcd{k}\opc{k} = 1-\opnq{k}$.
In this context, the parameters $t$ and $V$ may be interpreted as hopping
amplitude and density-density interaction strength respectively.
%
%
\subsection{Construction within the QISM framework}
The small polaron model can be associated to a graded six-vertex model with
anisotropy $\eta$ and $R$-matrix
\begin{equation}
 R(u)=\frac{1}{\sin(2\eta)}
      \begin{pmatrix}
        \sin(u+2\eta)&0&0&0\\
        0&\sin(u)&\sin(2\eta)&0\\
        0&\sin(2\eta)&\sin(u)&0\\
        0&0&0&-\sin(u+2\eta)
      \end{pmatrix}\ .
 \label{eq:RMatrix}
\end{equation}
$R(u)$ a solution to the Yang-Baxter Equation (YBE)
\begin{equation}
  R_{12}(u-v) R_{13}(u) R_{23}(v) = R_{23}(v) R_{13}(u) R_{12}(u-v)
  \label{eqAP:YBE}
\end{equation}
and enjoys several useful properties, such as
\begin{subequations}
\begin{itemize}
 \item \herv{P-symmetry}
       \begin{equation}
         R_{21}(u) \equiv \mathcal{P}_{12} R_{12}(u) \mathcal{P}_{12} = R_{12}(u)
         \label{eq:RMatrixPropertiesPSymmetry}
       \end{equation}
 \item \herv{T-symmetry}
       \begin{equation}
         R^{\st_1 \st_2}_{12}(u) = R^{\ist_1 \ist_2}_{12}(u) = R_{21}(u)
          \label{eq:RMatrixPropertiesTSymmetry}
       \end{equation}
 \item \herv{regularity}
       \begin{equation}
         R_{12}(0) = \mathcal{P}_{12}
         \label{eq:RMatrixPropertiesRegularity}
       \end{equation}
 \item \herv{unitarity}
       \begin{equation}
         R_{12}(u) R_{21}(-u) = \zeta(u)
          \label{eq:RMatrixPropertiesUnitarity}
       \end{equation}
       where the scalar function $\zeta(u)$ is given by
       \begin{equation*}
        \zeta(u)\equiv g(u) g(-u)\qquad\text{and}\qquad
        g(u)\equiv-\frac{\sin(u-2\eta)}{\sin(2\eta)}\;. 
       \end{equation*}
       Unitarity of an $R$-matrix is of course a direct consequence of its
       regularity. 
 \item \herv{crossing symmetry}
       \begin{equation}
         R_{21}^{\st_2}(-u-4\eta)R_{21}^{\st_1}(u) = \zeta(u+2\eta)
         \label{eq:RMatrixPropertiesCrossingSymmetry}
       \end{equation}
 \item \herv{periodicity}
       \begin{equation}
        R_{12}(u+\pi) = -\sigma^z_2\ R_{12}(u)\ \sigma^z_2 = -\sigma^z_1\
        R_{12}(u)\ \sigma^z_1 
        \label{eq:RMatrixPropertiesPeriodicity}
       \end{equation}
       The periodicity $R(u+2\pi)=R(u)$ is obvious from definition
       (\ref{eq:RMatrix}). 
\end{itemize}
\label{eq:RMatrixProperties}
\end{subequations}
The operations of partial super transposition $\st_a$ and inverse partial
super transposition $\ist_a$ as well as the graded permutation operator
$\mathcal{P}_{ab}$ and the notion of super tensor product structures are
explained in appendix \ref{app:GradedVectorSpaces}. Unless stated otherwise,
all embeddings are to be understood in a \herv{graded} sense, that is into a
super tensor product structure. Considering the Yang-Baxter Algebra (YBA) 
\begin{equation}
	R_{12}(u-v)\ T_1(u)\ T_2(v) = T_2(v)\ T_1(u)\ R_{12}(u-v)
	\label{eq:YBA}
\end{equation}
this means that $T_1(u)\equiv T(u)\sTP\eins$ and $T_2(v)\equiv \eins\sTP T(v)$.

The small polaron model constructed here is fundamental, i.e. the
Lax-operators $L_j(u)$, being local solutions to (\ref{eq:YBA}) are just
graded embeddings of the above $R$-Matrix (\ref{eq:RMatrix}), 
\begin{equation}
 L_j(u)=\frac{1}{\sin(2\eta)}
        \begin{pmatrix}
          \sin(u)\opn{j}+\sin(u+2\eta)\opnq{j}&\sin(2\eta)\opcd{j}\\
          \sin(2\eta)\opc{j}&\sin(u)\opnq{j}-\sin(u+2\eta)\opn{j}
        \end{pmatrix}\, .
 \label{eq:Lax}
\end{equation}
As a consequence of the YBA's co-multiplication property, a specific global
representation, the so-called monodromy matrix, can be constructed as
a product of Lax-operators taken in auxiliary space,
\begin{equation}
	T(u) \equiv L_N(u)\cdot\ldots\cdot L_2(u)\cdot L_1(u)
\end{equation} 
and gives rise to a family of commuting (super) transfer matrices 
\begin{equation}
	\tau(u) \equiv \str{T(u)} \quad\Rightarrow\quad \eckklammer{\tau(u) ,
          \tau(v)} = 0 \quad\forall u,v\in\mathbb{C}\ , 
\end{equation}
where $\str{\cdot}$ denotes the supertrace defined in appendix
\ref{app:GradedVectorSpaces}. In particular, the PBC hamiltonian
(\ref{eq:PBCHamiltonian}) with $t=1$ and $V=-\cos(2\eta)$ is among these
commuting operators, 
\begin{equation}
	H^\sss{PBC} = \left. -\sin(2\eta)
          \frac{\operatorname{d}}{\operatorname{d}\! u} \operatorname{ln}
          \tau(u) \right|_{u=0}\ . 
\end{equation}
%
\subsection{Asymptotic behaviour of the PBC transfer matrix}
By construction the monodromy matrix (and similarly the transfer matrix) is a
Laurent polynomial in $z\equiv\mathrm{e}^{\mathrm{i} u}$, i.e. $T(u) =
\sum_{k=-N}^{N} T_k z^k$.
For $z\to\infty$ the Lax-operators (\ref{eq:Lax}) are
\begin{equation}
	L_j(u)\approx \frac{z}{2\mathrm{i} \sin(2\eta)}
	              \begin{pmatrix}
		            \opn{j}+\mathrm{e}^{2\mathrm{i}\eta}\ \opnq{j} & 0 \\
		            0 & \opnq{j}-\mathrm{e}^{2\mathrm{i}\eta}\ \opn{j}
		          \end{pmatrix}
	\label{eq:AsymptoticLax_PBC}
\end{equation}
and consequently the asymptotic behaviour of the (super) transfer matrix is
given by 
\begin{equation}
  \tau(u) \approx \klammer{\frac{z}{2\mathrm{i} \sin(2\eta)}}^N \mathrm{e}^{\mathrm{i} N \eta}
  \eckklammer{
    \prod_{j=1}^N \klammer{\mathrm{e}^{-\mathrm{i} \eta}\opn{j} +
      \mathrm{e}^{\mathrm{i} \eta}\opnq{j}} - 
    \prod_{j=1}^N \klammer{\mathrm{e}^{-\mathrm{i} \eta}\opnq{j} -
      \mathrm{e}^{\mathrm{i} \eta}\opn{j}} 
  }\ . 
\end{equation}
As the leading term comprises only diagonal operators the first order
contributions to the transfer matrix eigenvalues $\Lambda_M(u)$ can easily be
determined and are found to depend on the total number of particles $M$,
\begin{equation}
  \Lambda_M(u) \approx \mathrm{e}^{\mathrm{i} u N}\klammer{\frac{\mathrm{e}^{\mathrm{i} \eta}}{ \mathrm{e}^{\mathrm{i}\,
        2\eta}-\mathrm{e}^{-\mathrm{i}\, 2\eta} }}^N 
  \klammer{
    \mathrm{e}^{\mathrm{i} N \eta} \mathrm{e}^{-\mathrm{i} M 2\eta} -
    (-1)^M \mathrm{e}^{-\mathrm{i} N \eta} \mathrm{e}^{\mathrm{i} M 2\eta} 
  }\ .
  \label{eq:Asmptotics_PBC}
\end{equation}
This result will be used to fix the degree of the $Q$-functions in section
\ref{sec:TQ_PBC}.

\section{Fusion of the $R$-matrix in auxiliary space}
Given an R-matrix as solution to the YBE (\ref{eqAP:YBE}), the fusion
procedure \cite{KuRS81,MeNe92a,Zhou96b} allows for the construction of
\herv{larger} R-matrices as solutions to corresponding Yang-Baxter equations,
where larger refers to the dimensionality of the auxiliary space involved. All
that fusion requires is a pair of complementary orthogonal\footnote{As usual,
  \herv{orthogonal} means $P_{12}^+\ P_{12}^- = 0$ whereas
  \herv{complementary} refers to the property $P_{12}^+ + P_{12}^- = \eins$.}
projectors $P_{12}^+$ and $P_{12}^-$ such that for a specific value of
$\rho\in\mathbb{C}$ the following \herv{triangularity condition} holds for
arbitrary spectral parameters $u\in\mathbb{C}$,
\begin{equation}
  P_{12}^-\ R_{13}^{\phantom{+}}(u)\ R_{23}^{\phantom{+}}(u+\rho)\ P_{12}^+ = 0\ .
  \label{eq:GenericRTC}
\end{equation} 
By virtue of this condition, it can be shown that the \herv{fused} R-Matrix,
defined by 
\begin{equation}
  R_{(12)3}(u) \equiv P^+_{12} R_{13}(u) R_{23}(u+\rho) P^+_{12}\,.
  \label{eq:fusedRMatrix}
\end{equation}
satisfies the corresponding Yang-Baxter equation
\begin{equation}
  R_{(12)3}(u-v)\ R_{(12) 4}(u)\ R_{34}(v) = R_{34}(v)\ R_{(12)4}(u)\ R_{(12) 3}(u-v)\ .
\end{equation}

It is easily found that the small polaron R-matrix (\ref{eq:RMatrix}) has two
distinct singularities at $u=\pm 2\eta$, 
\begin{equation}
  \det\{R(u)\}=-\frac{\sin(u-2\eta)}{\sin(2\eta)}
  \klammer{\frac{\sin(u+2\eta)}{\sin(2\eta)}}^3\stackrel{!}{=}0\,.  
\end{equation}
At $u=-2\eta$ the R-Matrix gives rise to a projector onto a one-dimensional
subspace, 
\begin{equation}
  P^- \equiv -\frac{1}{2}R(-2\eta) = \frac{1}{2}\begin{pmatrix}
                                       0 &  0 &  0 & 0 \\
                                       0 &  1 & -1 & 0 \\
                                       0 & -1 &  1 & 0 \\
                                       0 &  0 &  0 & 0 \\
                                     \end{pmatrix}\,.
\end{equation}
However, unlike in the case of the Heisenberg spin chain, the orthogonal
projector $P^+$ onto the complementary three-dimensional subspace cannot be
obtained from the R-matrix at the second singularity,
\begin{equation}
  P^+ \equiv \eins -P^- 
 \neq \frac{1}{2}R(2\eta)\,. 
\end{equation}
Using this projector, fusion of two small polaron R-matrices in the auxiliary
space can be achieved by means of (\ref{eq:fusedRMatrix}) with $\rho=2\eta$,  
\begin{equation}
  R_{(12)3}(u) \equiv P^+_{12} R_{13}(u) R_{23}(u+2\eta) P^+_{12}\,.
\end{equation}
The resulting object $R_{(12)3}(u)$ is an $8\!\times\! 8$-matrix of rank $6$
and may therefore be effectively reduced to a $6\!\times\! 6$-matrix $R_{\ll
  12\gg 3}(u)$ acting on a three-dimensional auxiliary space $V_{\ll 12\gg}$
and on a two-dimensional quantum space $V_3$. 
Changing from the $BFFB$-graded\footnote{This notation is
  explained in appendix \ref{app:GradedVectorSpaces}.} canonical basis
\begin{equation}
  \mathcal{B}_0 = \{e_1,\, e_2,\, e_3,\, e_4 \}_\sss{BFFB} \equiv
  \{\ket{0}\otimes\ket{0},\,\ket{0}\otimes\ket{1},\, \ket{1}\otimes\ket{0},\,
  \ket{1}\otimes\ket{1} \}_\sss{BFFB} 
\end{equation}
to the projectors' $BFBF$-graded singlet/triplet-eigenbasis
\begin{equation}
  \mathcal{B}_{\pm} = \{f_1,\, f_2,\, f_3,\, f_4 \}_\sss{BFBF} \equiv \{
  e_1,\, \frac{e_2+e_3}{\sqrt{2}},\, e_4,\, \frac{e_2-e_3}{\sqrt{2}}
  \}_\sss{BFBF}\, . 
\end{equation}
%
%
%
the matrix $R_{(12)3}(u)$ gains the advantageous shape
\begin{equation}
  \begin{pmatrix}
    R_{\ll 12\gg 3}(u) & \vline & ~ \\ \hline
                     ~ & \vline & \begin{matrix} 0 & 0 \\ 0 & 0 \end{matrix} 
  \end{pmatrix}^i_{~j}
  = (f_i)^T\eckklammer{R_{(12)3}(u)}f_j
\end{equation}
where $R_{\ll 12\gg 3}(u)$ is the only non-vanishing block. Explicitly one finds,
\begin{equation}
  \begin{split}
    R_{\ll 12\gg 3}(u) \propto
    \begin{pmatrix}
      \scriptstyle 2\sin(u+4\eta) & \scriptstyle 0
      & \vline & \scriptstyle 0                   \hspace{-1em}
      & \scriptstyle 0                   & \vline & \scriptstyle
      0                      & \scriptstyle 0 \\ 
      \scriptstyle 0              & \scriptstyle 2\sin(u)
      & \vline & \scriptstyle \sqrt{2}\sin(4\eta) \hspace{-1em}
      & \scriptstyle 0                   & \vline & \scriptstyle
      0                      & \scriptstyle 0 \\ \hline	 
      \scriptstyle 0              & \scriptstyle
      2\sqrt{2}\sin(2\eta) & \vline & \scriptstyle
      2\sin(u+2\eta)      \hspace{-1em} & \scriptstyle 0
      & \vline & \scriptstyle 0                      &
      \scriptstyle 0 \\ 
      \scriptstyle 0              & \scriptstyle 0
      & \vline & \scriptstyle 0                   \hspace{-1em}
      & \scriptstyle -2\sin(u+2\eta)     & \vline & \scriptstyle
      -2\sqrt{2}\sin(2\eta)  & \scriptstyle 0 \\ \hline 
      \scriptstyle 0              & \scriptstyle 0
      & \vline & \scriptstyle 0                   \hspace{-1em}
      & \scriptstyle \sqrt{2}\sin(4\eta) & \vline & \scriptstyle
      2\sin(u)               & \scriptstyle 0 \\ 
      \scriptstyle 0              & \scriptstyle 0
      & \vline & \scriptstyle 0                   \hspace{-1em}
      & \scriptstyle 0                   & \vline & \scriptstyle
      0                      & \scriptstyle 2\sin(u+4\eta) 
    \end{pmatrix}\, .
  \end{split}
\end{equation}
%

\subsection{General construction of higher fused $R$-matrices}
In general, higher fused R-matrices can be constructed employing the
projection operators 
\begin{equation}
  P^{+}_{1\ldots n} \equiv \frac{1}{n!}\sum_{\sigma\in \mathcal{S}_n}P_\sigma\ .
  \label{eq:SU2Projectors}
\end{equation}
Here $\sigma$ runs through all the elements of the permutation group
$\mathcal{S}_n$ and $P_\sigma$ is the permutation operator corresponding to
$\sigma$. 
Now the higher fused R-matrices are obtained as
\begin{equation}
  R_{(1\ldots n) q}(u) \equiv P^{+}_{1\ldots n}\ R_{1q}(u)\ R_{2q}(u+2\eta)\
  \ldots\ R_{nq}(u+[n-1]\cdot 2\eta) P^{+}_{1\ldots n} 
\end{equation}
Just as for the first fusion step, it is convenient to apply a similarity
transformation $A_{(1\ldots n)}$ into the eigenbasis\footnote{Since the
  projectors here are just the same as for the XXZ Heisenberg spin chain, the
  respective transformation is simply given by the matrix of Clebsch-Gordan
  coefficients.} of the projection operators, 
\begin{equation}
  A_{(1\ldots n)} R_{(1\ldots n)q}(u) A^{-1}_{(1\ldots n)} 
  \equiv \begin{pmatrix}
    R_{\ll 1\ldots n\gg q}(u) & \vline & \begin{matrix} ~ & ~ \\
      ~ & ~      \end{matrix} & ~ \\ \hline 
    ~  & \vline & \begin{matrix} 0 & ~ \\
      ~ & \ddots \end{matrix} & ~ 
  \end{pmatrix}\ .
  \label{eq:HigherFusedRMatrices}
\end{equation}
The first few ($n=1,2,3,4$) transformation matrices $A_{(1\ldots n)}$ are
explicitly given in appendix \ref{app:TrafoMatrices}.  By construction, all
matrix elements of (\ref{eq:HigherFusedRMatrices}), except for those in the
upper left $2(n+1)\times 2(n+1)$ block, vanish.  
%
This block is referred to as the fused R-matrix $R_{\ll 1\ldots n\gg q}(u)$.
As shown in table \ref{tb:AuxSpaceGrading}, its
fused auxiliary space has alternating gradation (bosonic, fermionic, \ldots).

\begin{table}[ht]\center
  \begin{tabular}[t]{rcccccc}
    \toprule[1.5pt]
    auxiliary space: &&  ~ $\ll\! 12\!\gg$ ~ & ~ $\ll\! 123\!\gg$ ~ & ~ $\ll\!
    1234\!\gg$ ~ & ~ $\ll\! 12345\!\gg$ ~ & ~ \ldots\\ 
    \midrule
    grading:         && $BFB$                & $BFBF$               & $BFBFB$
    & $BFBFBF$               & ~ \ldots\\ 
    \bottomrule[1.5pt]
  \end{tabular}
  \caption{Gradation of the fused auxiliary spaces in the projector eigenbasis.}
  \label{tb:AuxSpaceGrading}
\end{table}

The periodicity property (\ref{eq:RMatrixPropertiesPeriodicity}) carries over
to the fused $R$-matrices, 
\begin{equation}
  R_{\ll 1\ldots n\gg q}(u+\pi) = (-1)^n\ \sigma^z_{\ll n\gg}\ R_{\ll 1\ldots
    n\gg q}(u)\ \sigma^z_{\ll n\gg} 
  \label{eq:fusedRMatrixPropertiesPeriodicity}
\end{equation}
with $\sigma^z_{\ll n\gg}$ being defined through
\begin{equation}
	\sigma_{(n)}^z \equiv \prod_{k=1}^n \sigma_k^z\ \qquad
        \text{and}\qquad A_{(12\ldots n)}\ \sigma_{(n)}^z\ A^{-1}_{(12\ldots
          n)} 
	\equiv \begin{pmatrix}
               \sigma_{\ll n\gg}^z & \vline & \begin{matrix} ~ & ~ \\ ~ & ~      \end{matrix} & ~ \\ \hline
                                ~  & \vline & \begin{matrix} * & ~ \\ ~ & \ddots \end{matrix} & ~
           \end{pmatrix}\ .
 \label{eq:fusedSigmaZ}
\end{equation}

\subsection{Fusion hierarchy for super transfer matrices}

Since, by construction, the fused $R$-matrices again satisfy the YBE they can
be used to establish further families of commuting operators as supertraces of
fused monodromy matrices, 
\begin{eqnarray}
	T_{(1 2 \ldots n)}(u) &\equiv& P^{\sss{+}}_{1 2 \ldots n}\ R_{(1 2
          \ldots n) q_N}(u)\ \cdot\ldots\cdot\ R_{(1 2 \ldots n) q_2}(u)\
        R_{(1 2 \ldots n) q_1}(u)\ 
	                               P^{\sss{+}}_{1 2 \ldots n}\notag \\
	                      &=&      P^{\sss{+}}_{1 2 \ldots n}\ T_{(1 2
                                \ldots n-1)}(u)\ T_{n}(u+[n-1]\cdot 2\eta)\
                              P^{\sss{+}}_{1 2 \ldots n} \,,
\end{eqnarray}
\begin{equation}
  A_{(1 2\ldots n)}\ T_{(1 2 \ldots n)}(u)\ A^{-1}_{(1 2\ldots n)}
  \equiv \begin{pmatrix}
    T_{\ll 1 2\ldots n\gg}(u) & \vline & \begin{matrix} ~ & ~
      \\ ~ & ~      \end{matrix} & ~ \\ \hline 
    ~  & \vline & \begin{matrix} 0 & ~
      \\ ~ & \ddots \end{matrix} & ~ 
  \end{pmatrix}\ .
\end{equation}
Indeed it is found that the (super) transfer matrices obtained from any fusion
level $n$, 
\begin{equation}
  \tau^\sss{(n)}(u) \equiv \sTr{(1 2 \ldots n+1)}{T_{(1 2 \ldots
      n+1)}(u)} = \sTr{\ll 1 2 \ldots n+1\gg}{T_{\ll 1 2 \ldots
      n+1\gg}(u)}\ , 
\end{equation}
commute with the transfer matrices of any other fusion level $m$,
i.e. $\eckklammer{\tau^\sss{(n)}(u) , \tau^\sss{(m)}(v)} = 0$ for all
$u,v\in\mathbb{C}$ and arbitrary $n,m\in\mathbb{N}_0$.  
A most interesting fact is, that these \herv{fused} transfer matrices obey
certain functional relations, known as \herv{fusion hierarchy}. For the
periodic boundary case, the fusion hierarchy reads, 
\begin{equation}
\label{eq:FusionHierarchyPBC}
  \tau^\sss{(n)}(u)\
  \tau^\sss{(0)}(u+[n+1]\cdot2\eta)=\tau^\sss{(n+1)}(u)+\delta(u+n\cdot
  2\eta)\tau^\sss{(n-1)}(u)\ , 
\end{equation}
where $\delta(u)\equiv\delta\gklammer{T(u)}$ labels the PBC super quantum
determinant (SQD) defined in appendix \ref{app:SQD}. In contrast to ungraded
models, such as the XXZ Heisenberg spin chain, this quantum determinant is
\herv{not} proportional to the identity. 
%
\section{TQ-equations for PBC}
\label{sec:TQ_PBC}
After applying a shift $u\rightarrow u-[n+1]\cdot 2\eta$ the PBC fusion
hierarchy (\ref{eq:FusionHierarchyPBC}) reads 
\begin{equation}
 \tau^{(n)}(u-[n+1]\cdot 2\eta)\ \tau^{(0)}(u) = \tau^{(n+1)}(u-[n+1]\cdot
 2\eta) + \delta(u-2\eta)\ \tau^{(n-1)}(u-[n+1]\cdot 2\eta)\ . 
 \label{eq:ShiftedFusionHierarchyPBC}
\end{equation}
As all operators in this equation mutually commute, it may equally well be read as an equation for the eigenvalues $\Lambda^{(n)}(u)$ of the fused super transfer matrices. With $\Lambda(u)\equiv\Lambda^{(0)}(u)$ this yields
\begin{equation}
	\Lambda(u)=\frac{\Lambda^{(n+1)}(u-[n+1]\cdot 2\eta)}{\Lambda^{(n)}(u-[n+1]\cdot 2\eta)}
	           -(-1)^{N+M} \zeta^N(u) \frac{\Lambda^{(n-1)}(u-[n+1]\cdot 2\eta)}{\Lambda^{(n)}(u-[n+1]\cdot 2\eta)}
\end{equation}
where $M$ is the number of particles in the system, such that the sign $(-1)^M$ depends on the parity of the corresponding eigenstate (bosonic/fermionic). This pecularity stems from the fact, that the PBC SQD (\ref{eqAP:SQD}) can not simply be treated as a scalar function but rather as an operator that intersperses sign factors into the respective sectors.
This may be illustrated by considering the fusion hierarchy (\ref{eq:ShiftedFusionHierarchyPBC}) in a diagonal basis for chain length $N=1$,
\begin{equation}
	\begin{pmatrix}
		* & ~ \\
		~ & *  
    \end{pmatrix}
    \begin{pmatrix}
		* & ~ \\
		~ & * 
    \end{pmatrix}
    =
    \begin{pmatrix}
		* & ~ \\
		~ & * 
    \end{pmatrix}
    +
    \begin{pmatrix}
		+ & ~ \\
		~ & -
    \end{pmatrix}
   	\begin{pmatrix}
		* & ~ \\
		~ & *
    \end{pmatrix}\quad
   \begin{matrix}
	\leftarrow\ B\\ \leftarrow\ F
   \end{matrix}
\end{equation}
Introducing the functions
\begin{equation}
	\bar{Q}^{(n)}(u) \equiv \Lambda^{(n)}(u-[n+1]\cdot 2\eta)
\end{equation}
the eigenvalues can be rewritten as
\begin{equation}
	\Lambda(u)=\frac{ \bar{Q}^{(n+1)}(u+2\eta) }{ \bar{Q}^{(n)}(u) }
	           -(-1)^{N+M} \zeta^N(u) \frac{ \bar{Q}^{(n-1)}(u-2\eta) }{ \bar{Q}^{(n)}(u) }\ .
\end{equation}
Now factorize $\bar{Q}^{(n)}$ according to
\begin{equation}
	\bar{Q}^{(n)}=\chi_M(u)\cdot \Upsilon_n^N(u)\cdot Q^{(n)}(u)
\end{equation}
where
\begin{equation}
	\chi_M(u)  \equiv \mathrm{e}^{\mathrm{i}\pi(M+1)\frac{u}{2\eta}}
	\quad\text{and}\quad
    \Upsilon_n(u) \equiv \prod_{k=0}^n \frac{\sin(u-[n-k+1]\cdot 2\eta)}{\sin(2\eta)}\ .
\end{equation}
%
%
%
Assuming the existence of the limit $Q(u)\equiv \lim_{n\rightarrow\infty}
Q^{(n)}(u)$, this yields 
\begin{equation}
	\Lambda(u) = \klammer{\frac{\sin(u+2\eta)}{\sin(2\eta)}}^N \frac{ Q(u-2\eta) }{ Q(u) }
                 -(-1)^M \klammer{\frac{\sin(u)}{\sin(2\eta)}}^N \frac{ Q(u+2\eta) }{ Q(u) }\ .
    \label{eq:TQEquation_PBC}
\end{equation}
Due to the structure of the entries in the Lax-operators, the $Q$-functions
factorize 
%
\begin{equation}
	Q(u) = \prod_{\ell=1}^G \sin(u-\lambda_\ell)
	\label{eq:QAnsatz_PBC}
\end{equation}
where the integer $G$ can be determined by considering the asymptotic
behaviour of $\Lambda(u)$. In the limit $z\equiv \mathrm{e}^{\mathrm{i} u}\rightarrow
\infty$ the leading contribution to (\ref{eq:TQEquation_PBC}) is
\begin{equation}
  \Lambda(u) \approx \mathrm{e}^{\mathrm{i} N u} \klammer{\frac{\mathrm{e}^{\mathrm{i} \eta}}{\mathrm{e}^{\mathrm{i}\,
        2\eta}-\mathrm{e}^{-\mathrm{i}\, 2\eta}}}^N 
  \eckklammer{ \mathrm{e}^{\mathrm{i} N \eta}\ \mathrm{e}^{-\mathrm{i} G\, 2 \eta} - (-1)^M \mathrm{e}^{-\mathrm{i} N
      \eta}\ \mathrm{e}^{\mathrm{i} G\, 2 \eta} } 
\end{equation}
such that consistency with (\ref{eq:Asmptotics_PBC}) immediately fixes $G =
M$. The requirement for the eigenvalues $\Lambda(u)$ to be analytic ultimately
yields 
\begin{equation}
  \operatorname{Res}_{\lambda_j}(\Lambda) = 0
  \quad\Leftrightarrow\quad
  \klammer{\frac{\sin(\lambda_j+2\eta)}{\sin(\lambda_j)}}^N = 
  \prod_{\ell=1}^M\frac{\sin(\lambda_j-\lambda_\ell+2\eta)}{\sin(\lambda_\ell
    -\lambda_j+2\eta)}  
\end{equation}
which are precisely the Bethe equations for this model
\cite{PuZh86,ZhJW89,UmFW99}.  Compared to the periodic XXZ Heisenberg chain
these Bethe equations exhibit an additional sign, reflecting the different
twist in the boundary conditions appearing in the sectors with even and odd
particle numbers through the Jordan-Wigner transformation from the fermionic
to the spin model.

\section{Truncation of the PBC fusion hierarchy}
In the case of the XXZ-model it has been observed, that for certain values of
the anisotropy $\eta$ the fusion hierarchy repeats itself after a
finite number of steps. The small polaron model shares this feature at values
$\eta=\eta_p$ where
\begin{equation}
  \eta_p \equiv \frac{\pi/2}{p+1}\ .
\end{equation}

\subsection{$R$-matrix truncation}
The truncation identities for the R-matrices are found to be
\begin{equation}
  \mathcal{R}^{(p)}_{q}(u,\eta_p)
  = \begin{pmatrix}
       -\mathcal{M}_p(u)\ \sigma^z_q & ~ & ~ \\
        ~  & \zeta(u)\sigma^z_{q}\ \mathcal{R}^{(p-2)}_{q}(u+2\eta_p,\eta_p) & ~ \\
        ~  & ~ & \mathcal{M}_p(u)\ (\sigma^z_q)^p
    \end{pmatrix}
  \label{eq:RMatrixTruncation}
\end{equation}
where
\begin{equation}
 \begin{split}
  \mathcal{R}^{(p)}_{q}(u,\eta) &\equiv 	B_{\ll 1\ldots(p+1)\gg}\ R_{\ll 1\ldots(p+1)\gg q}(u)\ B^{-1}_{\ll 1\ldots(p+1)\gg}\\
  \mathcal{M}_p(u) &\equiv \klammer{\frac{1/2}{\sin(2\ \eta_p)}}^p\ \frac{\sin([p+1]\ u)}{\sin(2\ \eta_p)}
 \end{split}
 \label{eq:RMatrixTruncationDefs}
\end{equation}
with the transformation matrices $B_{\ll 1\ldots n\gg}$ explicitly given in appendix \ref{app:TrafoMatrices} up to $n=4$.

\subsection{Super transfer matrix truncation}
\label{ss:SuperTransfermatrixTruncation}
For periodic boundary conditions the $B$-transformed fused monodromy matrix
$\mathcal{T}^{(p)}(u,\eta)$ of an $N$-site model with quantum space
$\mathcal{H}=V_{q_1}\sTP V_{q_2} \sTP \ldots \sTP V_{q_N} $ is defined as 
\begin{eqnarray}
  \mathcal{T}^{(p)}(u,\eta) &\equiv& \mathcal{R}^{(p)}_{q_N}(u,\eta)\
  \mathcal{R}^{(p)}_{q_{N-1}}(u,\eta)\ \ldots \mathcal{R}^{(p)}_{q_1}(u,\eta)\
  \\ 
    &=& B_{\ll 1\ldots (p+1)\gg}\ R_{\ll 1\ldots (p+1)\gg q_N}(u)\ \ldots\
    R_{\ll 1\ldots (p+1)\gg q_1}(u)\ B^{-1}_{\ll 1\ldots (p+1)\gg} \notag \\ 
    &=& B_{\ll 1\ldots (p+1)\gg}\ T_{\ll 1\ldots (p+1)\gg}(u)\ B^{-1}_{\ll
      1\ldots (p+1)\gg} \notag  
\end{eqnarray}
and due to the cyclic invariance of the supertrace it yields the exact same transfer matrix
\begin{equation}
  \tau^{(p)}(u,\eta) \equiv \sTr{\ll 1\ldots (p+1)\gg}{T_{\ll 1\ldots (p+1)\gg}(u)} = \sTr{\ll 1\ldots (p+1)\gg}{\mathcal{T}^{(p)}(u,\eta)}\ .
\end{equation} 
At $\eta = \eta_p$ the truncation identity (\ref{eq:RMatrixTruncation}) for
R-matrices gives 
\begin{equation}
\begin{aligned}
  &\hspace{-1em}\mathcal{T}^{(p)}(u,\eta_p) = \\
  &\begin{pmatrix}
      [-\mathcal{M}_p(u)]^N \prod_{i=N}^1 \sigma^z_{q_i} & ~ & ~ \\[.5em]
      ~ & \zeta^N(u) \prod_{i=N}^1 \sigma^z_{q_i}\
      \mathcal{R}^{(p-2)}_{q_i}(u+2\eta_p,\eta_p)  & ~ \\[.5em] 
      ~ & ~ & [\mathcal{M}_p(u)]^N \prod_{i=N}^1 (\sigma^z_{q_i})^p
    \end{pmatrix}
  \end{aligned}
\end{equation}
such that the truncation identity for the transfer matrices is found to be
\begin{equation}
  \begin{aligned}
    \tau^{(p)}(u,\eta_p) &= [-\mathcal{M}_p(u)]^N\ \klammer{\prod_{i=1}^N
      \sigma^z_{q_i}} - (-1)^{p} [\mathcal{M}_p(u)]^N\ \klammer{\prod_{i=1}^N
      (\sigma^z_{q_i})^p}  \\ 
    &~ - \zeta^N(u) \klammer{\prod_{i=1}^N \sigma^z_{q_i}}
    \tau^{(p-2)}(u+2\eta_p,\eta_p)\ . 
  \end{aligned}
\end{equation}

\section{The small polaron with open boundary conditions}
%
\subsection{Reflection algebras and boundary matrices}
\label{ch:ReflectionAlgebrasAndBoundaryMatrices}
The construction of integrable systems with open boundary conditions is based
on representations of the graded reflection algebra
\begin{equation}
\begin{split}
  R_{12}(u-v)\calT^{-}_{1}(u) & R_{21}(u+v)\calT^{-}_{2}(v)\\[.5em]
          =\,\calT^{-}_{2}(v) & R_{12}(u+v)\calT^{-}_{1}(u)R_{21}(u-v)
\end{split}
  \label{eq:RA1}
\end{equation}
and the corresponding dual graded reflection algebra
\begin{equation}
\begin{split}
  \bar{R}_{12}(v-u)\calT^{+}_{1}(u)^{\sT{1}} & R_{21}(-u-v-4\eta)\calT^{+}_{2}(v)^{\isT{2}}\\[.5em]
                         =\,\calT^{+}_{2}(v)^{\isT{2}} & R_{12}(-u-v-4\eta)\calT^{+}_{1}(u)^{\sT{1}}\bar{R}_{21}(v-u) .
\end{split}
  \label{eq:RA2}
\end{equation}
The relation between $R_{ab}(u)$ and the \herv{conjugated $R$-matrix}
$\bar{R}_{ab}(u)$ is explained in appendix \ref{app:RelationToBracken}. 
$c$-number valued boundary matrices, compatible with the respective reflection
equation, are found to be \cite{Zhou96a,Zhou97,GuGR98} (see also
\cite{VeGo93} for the ungraded case of the XXZ chain)
\begin{equation}
 \begin{aligned}
   &K^\sss{-}(u)=\omega^\sss{-}
          \begin{pmatrix}
            \sin(u+\psi_\sss{-})&\alpha_\sss{-}\sin(2u)\\
            \beta_\sss{-}\sin(2u)&-\sin(u-\psi_\sss{-})
          \end{pmatrix}
\\ 
   &K^\sss{+}(u)=\omega^\sss{+}
          \begin{pmatrix}
            \sin(u+2\eta+\psi_\sss{+})&\alpha_\sss{+}\sin(2[u+2\eta])\\
            \beta_\sss{+}\sin(2[u+2\eta])&\sin(u+2\eta-\psi_\sss{+})
          \end{pmatrix}
\\
 \end{aligned}
 \label{eq:generalKs}
\end{equation}
with normalizations $\omega^\sss{\pm} \equiv \omega^\sss{\pm}(\eta)$ defined by
\begin{equation}
  \omega^\sss{-}(\eta) \equiv \frac{1}{\sin(\psi_\sss{-})} \qquad \text{and} \qquad
  \omega^\sss{+}(\eta) \equiv \frac{1}{2\cos(2\eta)\sin(\psi_\sss{+})} \, .	
\end{equation}
These matrices share the periodicity property of the $R$-matrix, i.e.
\begin{equation}
 K^\sss{\mp}(u+\pi) = -\sigma^z\ K^\sss{\mp}(u)\ \sigma^z\; .
 \label{eq:KPeriodicity}
\end{equation}
Here the normalizations were chosen such that
\begin{equation}
 K^\sss{-}(0)=\eins\qquad\text{and}\qquad\str{K^\sss{+}(0)}=1	\, ,
 \label{eq:KMatrixProperties}
\end{equation}
but apart from this, the two solutions are related via
\begin{equation}
 K^\sss{+}(u) = \left[ \frac{1}{2\cos(2\eta)} K^\sss{-}(-u-2\eta)\ \sigma^z_{\phantom{\xi_\xi}}\hspace*{-.3em}
 \right]_{\klammer{\ominus\ \rightarrow\ \oplus}}\ ,
 \label{eq:KpmRelation}
\end{equation}
where $\klammer{\ominus\ \rightarrow\ \oplus}$ marks the replacements
$\klammer{\alpha_\sss{-},\ \beta_\sss{-},\ \psi_\sss{-}}\ \rightarrow\
\klammer{-\alpha_\sss{+},\ \beta_\sss{+},\ -\psi_\sss{+}}$.  In principle the
parameters $\psi_\sss{\pm}$ are arbitrary even Grassmann numbers but their
invertability requires them to have a non-vanishing complex part\footnote{Such
  an additive part, that contains no nilpotent generators, is sometimes called
  the \herv{body} of a Grassmann number. It is to be distinguished from the
  \herv{soul} of a Grassmann number, which contains only sums of products of
  nilpotent generators.}. The remaining parameters $\alpha_\sss{\pm}$ and
$\beta_\sss{\pm}$ are odd Grassmann numbers, being subject to the condition
$\alpha_\sss{\pm}\cdot\beta_\sss{\pm}=0$. 

%

Given the monodromy matrix $T(u)=L_N(u)L_{N-1}(u)\dots L_1(u)$ it is possible
to construct a further representation of the reflection algebra (\ref{eq:RA1})
as
\begin{equation}
 \calT^-(u)=T(u)\ K^\sss{-}(u)\ \widehat{T}(u)\equiv\begin{pmatrix}\calA(u)&\calB(u)\\ \calC(u)&\calD(u)\end{pmatrix},
 \label{eq:RA_Representation}
\end{equation}
with $\widehat{T}(u)$ being a shorthand notation for $T^{-1}(-u)$, 
\begin{eqnarray}
  \widehat{T}(u) &\equiv& R^{-1}_{01}(-u)\ R^{-1}_{02}(-u)\ \ldots\ R^{-1}_{0N}(-u) \notag\\
                 &\stackrel{\sss{(\ref{eq:RMatrixPropertiesUnitarity})}}{=}& \frac{1}{\zeta^N(u)} R_{10}(u)\ R_{20}(u)\ \ldots\ R_{N0}(u)
                 \label{eq:MonodromyHat} \\
                 &\stackrel{\sss{(\ref{eq:RMatrixPropertiesPSymmetry})}}{=}& \klammer{\frac{1}{\zeta(u)}}^N R_{01}(u)\ R_{02}(u)\ \ldots\ R_{0N}(u)\ , \notag
\end{eqnarray}
resulting in an OBC super transfer matrix
\begin{equation}
	\tau(u)\equiv\sTr{0}{K^\sss{+}(u)\calT^-(u)}
	\label{eq:Transfermatrix_OBC}
\end{equation}
Expanding $\tau(u)$ around $u=0$ one obtains a Hamiltonian featuring the same
bulk part (\ref{eq:HamiltonianDensity}) as the corrsponding PBC Hamiltonian.
Defining the shorthands
$\mathcal{N}_\pm\equiv\frac{1}{2}\csc(2\eta)\csc(\psi_+)\sin(2\eta\pm\psi_+)$,
the resulting OBC Hamiltonian
\begin{equation}
	\begin{split}
	    H^{\text{\scriptsize{OBC}}}=\ &\sum_{j=1}^{N-1}H_{j,j+1} + \frac{1}{2}\cot(\psi_-)\eckklammer{\opnq{1}-\opn{1}}+ \eckklammer{\mathcal{N}_+\ \opnq{N}-\mathcal{N}_-\ \opn{N}}\\[.25em]
	                                  &+ \csc(\psi_-)\eckklammer{\alpha_-\ \opc{1}-\beta_-\ \opcd{1}} + \csc(\psi_+)\eckklammer{\alpha_+\ \opc{N}-\beta_+\ \opcd{N}}
    \end{split}
  \label{eq:OBCHamiltonian}
\end{equation}
is derived from the set of open boundary transfer matrices by
\begin{equation}
	\partial_u\  \tau(u) |_{u=0}=2\ H^{\text{\scriptsize{OBC}}}+\text{const.}\ .
\end{equation}
In the case of diagonal boundaries, i.e. $\alpha_\pm=\beta_\pm=0$, Bethe
equations can be derived using the algebraic Bethe ansatz.  This allows for
the computation of the transfer matrix eigenvalues and eigenvectors
(cf. appendix \ref{app:diagABA} resp. \cite{UmFW99}).  Here the eigenvalues
coincide with those of the spin-1/2 XXZ Heisenberg chain subject to (diagonal)
boundary magnetic fields.
%
\subsection{Properties of the OBC transfer matrix}
As a consequence of the properties
(\ref{eq:RMatrixPropertiesCrossingSymmetry}),
(\ref{eq:RMatrixPropertiesPeriodicity}) of the $R$-matrix and
(\ref{eq:KPeriodicity}), (\ref{eq:KpmRelation}) of the boundary matrices the
transfer matrix (\ref{eq:Transfermatrix_OBC}) of the small polaron model
enjoys several useful properties, such as
\begin{subequations}
\begin{itemize}
 \item \herv{$\pi$-periodicity}
       \begin{equation}
         \tau(u+\pi) = \tau(u)\,,
         \label{eq:TauPropertiesPiPeriodicity}
       \end{equation}
 	\item \herv{crossing symmetry}
	       \begin{equation}
	         \zeta^N(u)\ \tau(u)  = \zeta^N(-u-2\eta)\ \tau(-u-2\eta)\,.
	          \label{eq:TauPropertiesCrossing}
	       \end{equation}
\end{itemize}
\end{subequations}
In addition $\tau(u)$ is normalized as
\begin{equation}
  \tau(0)  = \eins
  \label{eq:TauPropertiesInitial}
\end{equation}
and becomes diagonal in the semi-classical limit $\eta\to0$:
\begin{equation}
  \begin{split}
    \left.\tau(u)\right|_{\eta=0} =\ &\frac{(-1)^N}{\sin(\psi_-)\sin(\psi_+)}
    \left( 2\sin^2(u)\cos^2(u)\ (\beta_+ \alpha_- - \alpha_+ \beta_-)\cdot \sigma^z_\sss{(N)} \phantom{\sum} \right. \\
    &\left. \phantom{\sum} - \eckklammer{\cos^2(u)\sin(\psi_-)\sin(\psi_+)+\sin^2(u)\cos(\psi_-)\cos(\psi_+)}\cdot \eins \right)\,.
  \end{split}
  \label{eq:TauPropertiesSemiClassical}
\end{equation}
The asymptotic behaviour of the (super) transfer matrix in the
limit $z\equiv \mathrm{e}^{\mathrm{i} u} \rightarrow \infty$ can be read off
from its construction: that of the Lax operators $L_j(u)$ is given in
Eq.~(\ref{eq:AsymptoticLax_PBC}).  Similarly we find
\begin{subequations}
  \begin{eqnarray}
    L^{-1}_j(-u) &=& \frac{4\sin(2\eta)}{2\mathrm{i}\, z}
    \begin{pmatrix}
      \opn{j}+\mathrm{e}^{2\mathrm{i}\eta}\ \opnq{j} & 0 \\
      0 & \opnq{j}-\mathrm{e}^{2\mathrm{i}\eta}\ \opn{j}
    \end{pmatrix} + \mathcal{O}\klammer{\frac{1}{z^2}}\,,
    \\
    K^\sss{-}(u) &=& \frac{\omega^\sss{-}}{2\mathrm{i}}
    \eckklammer{
      z^2\begin{pmatrix}
        0 & \alpha_- \\
        \beta_- & 0
      \end{pmatrix}
      +z\begin{pmatrix}
        \mathrm{e}^{\mathrm{i} \psi_-} & 0 \\
        0 & -\mathrm{e}^{-\mathrm{i}\psi_-}
      \end{pmatrix}
      +\mathcal{O}\klammer{\frac{1}{z}}
    }\,,
    \\
    K^\sss{+}(u) &=& \frac{\omega^\sss{+}\ \mathrm{e}^{2\mathrm{i} \eta}}{2\mathrm{i}}
    \eckklammer{
      z^2\mathrm{e}^{2\mathrm{i} \eta}\begin{pmatrix}
        0 & \alpha_+ \\
        \beta_+ & 0
      \end{pmatrix}
      + z\begin{pmatrix}
        \mathrm{e}^{\mathrm{i}\psi_+} & 0 \\
        0 & \mathrm{e}^{-\mathrm{i}\psi_+}
      \end{pmatrix}
      +\mathcal{O}\klammer{\frac{1}{z}}
    }\,.
  \end{eqnarray}
\end{subequations}
As a consequence, the asymptotics of the OBC transfer matrix
(\ref{eq:Transfermatrix_OBC}) and of their eigenvalues is given by
\begin{equation}
  \begin{split}
    \tau(u) &= (-1)^N\ \frac{\omega^\sss{+}\omega^\sss{-}}{4}\
    \mathrm{e}^{4\mathrm{i}\eta}\klammer{\beta_+\alpha_- - \alpha_+\beta_-} z^4 
    \prod_{j=1}^N( \opnq{j}-\mathrm{e}^{2\mathrm{i}\eta}\opn{j} )(
    \opn{j}+\mathrm{e}^{2\mathrm{i}\eta}\opnq{j} ) +\mathcal{O}(z^2)\\ 
    \Lambda_\pm(u) &= \pm(-1)^N\ \frac{\omega^\sss{+}\omega^\sss{-}}{4}\
    \mathrm{e}^{4\mathrm{i}\eta}\klammer{\beta_+\alpha_- - \alpha_+\beta_-} z^4
    \mathrm{e}^{\mathrm{i} N\, 2\eta} 
    +\mathcal{O}(z^2)\ .
  \end{split}
  \label{eq:NonDiagonalAsymptotics}
\end{equation}
The eigenvalues $\Lambda_\pm(u)$ have been classified according to a parity
which is determined by the (diagonal) operator controlling the asymptotics of
$\tau(u)$.

Note that in the case of diagonal boundaries, i.e.\ $\alpha_\pm = \beta_\pm =
0$, the $\mathcal{O}(z)$ terms of the $K$ matrices become the leading ones
such that
\begin{equation}
  \begin{split}
    \tau(u) &= -(-1)^N\ \frac{\omega^\sss{+}\omega^\sss{-}}{4}\ \mathrm{e}^{2\mathrm{i}\eta}\, z^2
    \left[
      \mathrm{e}^{\mathrm{i}(\psi_+ + \psi_-)} \prod_{j=1}^N( \opn{j}+\mathrm{e}^{2\mathrm{i}\eta}\opnq{j}
      )( \opn{j}+\mathrm{e}^{2\mathrm{i}\eta}\opnq{j} ) 
    \right. \\
    &\phantom{= (-1)^N\ \frac{-\mathrm{e}^{2\mathrm{i}\eta}\, z^2}{8\cos(2\eta)\sin(.)\sin(.)}}
    \left.
      ~+ \mathrm{e}^{-\mathrm{i}(\psi_+ + \psi_-)} \prod_{j=1}^N(
      \opnq{j}-\mathrm{e}^{2\mathrm{i}\eta}\opn{j} )( \opnq{j}-\mathrm{e}^{2\mathrm{i}\eta}\opn{j} ) 
    \right]+\mathcal{O}(z)\\
    \Lambda_M(u) &= -(-1)^N\ \frac{\omega^\sss{+}\omega^\sss{-}}{4}\
    \mathrm{e}^{2\mathrm{i}\eta}\, z^2 
    \klammer{
      \mathrm{e}^{\mathrm{i}(\psi_+ + \psi_-)}  \mathrm{e}^{4\mathrm{i} (N-M)\eta}
      +\mathrm{e}^{-\mathrm{i}(\psi_+ + \psi_-)} \mathrm{e}^{4\mathrm{i} M \eta}
    }
    +\mathcal{O}(z)\ .
  \end{split}
\end{equation}
Here, as in the case of periodic boundary conditions, the asymptotic behaviour
of the transfer matrix eigenvalues can be related to the (conserved) total
number $M$ of particles in the state.
%
\subsection{Fusion of the boundary matrices}
For the sake of readability it is convenient to define the following ordered product of $R$-matrices,
\begin{equation}
	R^\text{string}_i(u) \equiv \prod_{k=1}^i R_{k,i+1}(2u+[i+k-1]\cdot 2\eta)\ ,
\end{equation}
such that the fused $K^\sss{-}$ boundary matrices may be written as,
\begin{equation}
\label{eq:kmatm}
\begin{split}
	K^\sss{-}_{(12\ldots n)}(u) &\equiv P^\sss{+}_{12\ldots n} \eckklammer{
	                                       \prod_{i=1}^{n-1} K^\sss{-}_i(u+[i-1]\cdot 2\eta)\ R^\text{string}_i(u)
	                                     } K^\sss{-}_n(u+[n-1]\cdot 2\eta)\ P^\sss{+}_{12\ldots n}\\
	&\Rightarrow A_{(12\ldots n)}\ K^\sss{-}_{(12\ldots n)}(u)\ A^{-1}_{(12\ldots n)}
       \equiv \begin{pmatrix}
                K^\sss{-}_{\ll 1\ldots n\gg}(u) & \vline & \begin{matrix} ~ & ~ \\ ~ & ~      \end{matrix} & ~ \\ \hline
                                             ~  & \vline & \begin{matrix} 0 & ~ \\ ~ & \ddots \end{matrix} & ~
              \end{pmatrix}\ ,
\end{split}
\end{equation}
(see also Refs.~\cite{MeNe92a,Zhou96b,Nepo02} for the XXZ model)
where $K^\sss{-}_{(12\ldots n)}(u)$ is a $2^n\!\times\! 2^n$-matrix with $K^\sss{-}_{\ll 1\ldots n\gg}(u)$ being the only non-vanishing block of dimensions $(n+1)\times(n+1)$.
There is a useful relation between the fused $K^\sss{-}$- and $K^\sss{+}$-matrices that stems from (\ref{eq:KpmRelation}), 
\begin{eqnarray}
	K^\sss{+}_{(12\ldots n)}(u) &=& \left[\klammer{\frac{1}{2\cos(2\eta)}}^n
	                                   K^\sss{-}_{(n\ldots 21)}(-u-n\cdot 2\eta)\ \sigma^z_{(n)}
	\right]_{\klammer{\ominus\ \rightarrow\ \oplus}}\notag\\
	  &\Rightarrow& A_{(12\ldots n)}\ K^\sss{+}_{(12\ldots n)}(u)\ A^{-1}_{(12\ldots n)} \equiv
	           \begin{pmatrix}
                  K^\sss{+}_{\ll 1\ldots n\gg}(u) & \vline & \begin{matrix} ~ & ~ \\ ~ & ~      \end{matrix} & ~ \\ \hline
                                               ~  & \vline & \begin{matrix} 0 & ~ \\ ~ & \ddots \end{matrix} & ~
               \end{pmatrix}
\end{eqnarray}
%
and defines the $(n+1)\times(n+1)$-matrix $K^\sss{+}_{\ll 1\ldots n\gg}(u)$ in the obvious way, where $\sigma_{(n)}^z$ was defined in (\ref{eq:fusedSigmaZ}).
Note that the order of all spaces in $K^\sss{-}_{(n\ldots 21)}$ is
inverted. Thus, by changing the space labels according to $i \rightarrow
n+1-i$ the fused right boundary matrix may explicitly be written as 
\begin{eqnarray}
	K^\sss{+}_{(12\ldots n)}(u) &=& P^\sss{+}_{12\ldots n} \eckklammer{
	                                \prod_{i=1}^{n-1} K^\sss{+}_{n+1-i}(u+[n-i]\cdot 2\eta)\ \bar{R}^\text{string}_i(u)
	                              } K^\sss{+}_1(u)\ P^\sss{+}_{12\ldots n}\\
	\bar{R}^\text{string}_i(u) &\equiv& \prod_{k=1}^i \bar{R}_{n+1-k,n+1-(i+1)}(-2u+[i+k-1-2n]\cdot 2\eta)\ .
\end{eqnarray}
The reason why the conjugated $R$-matrices (\ref{eqAP:ConjugatedR}) appear in this expression is that by commuting the $\sigma^z$-matrices, arising from (\ref{eq:KpmRelation}), to the right, the relation
\begin{equation}
	\bar{R}_{ab}(u) = \sigma^z_a R_{ab}(u) \sigma^z_a = \sigma^z_b R_{ab}(u) \sigma^z_b
\end{equation}
is employed, cf. (\ref{eqAP:MConjugatedR}). 

Since $[P^\sss{+}_{(1\ldots n)}, \sigma_{(n)}^z] = 0$ and $[\sigma_{(n)}^z, A_{(12\ldots n)}\ \sigma_{(n)}^z\ A^{-1}_{(12\ldots n)}]=0$, the periodicity property (\ref{eq:KPeriodicity}) carries over to the fused $K^\sss{-}$-matrices,
\begin{subequations}
\begin{eqnarray}
	K^\sss{\mp}_{(12\ldots n)}(u+\pi) &=& (-1)^n\ \sigma_{(n)}^z K^\sss{\mp}_{(12\ldots n)}(u) \sigma_{(n)}^z\\
	K^\sss{\mp}_{\ll 12\ldots n\gg}(u+\pi) &=& (-1)^n\ \sigma_{\ll n\gg}^z K^\sss{\mp}_{\ll 12\ldots n\gg}(u) \sigma_{\ll n\gg}^z
\end{eqnarray}
\end{subequations}
where the alternating sign results from successive application of (\ref{eq:RMatrixPropertiesPeriodicity}).
%
\subsection{Fusion hierarchy for OBC}
From the fused quantities, it is again possible to derive a family of commuting operators
\begin{equation}
  \tau^{(n)}(u) \equiv \sTr{\ll 1\ldots n\gg}{
                        K^{\sss +}_{\ll 1\ldots n\gg}(u)\ T_{\ll 1\ldots n\gg}(u)\ K^{\sss -}_{\ll 1\ldots n\gg}(u)\ \widehat{T}_{\ll 1\ldots n\gg}(u+[n-1]\cdot 2\eta) 
                      }
  \label{eq:obcFusedTransfermatrix}
\end{equation}
that extends the existing familiy of commuting super transfer matrices $\tau(u)=\tau^{(1)}(u)$ such that $[ \tau^{(i)}(u) , \tau^{(k)}(v) ] = 0$ for all $i,j \geq 1$.
%
%
The quantity $\widehat{T}_{\ll 1\ldots n\gg}(u)$ appearing in (\ref{eq:obcFusedTransfermatrix}) is related to the fused object
\begin{eqnarray}
  \widehat{T}_{(1\ldots n)}(u+[n-1]\cdot 2\eta) &=& P_{1\ldots n}^{\sss +}\
                                                     \widehat{T}_1(u) \widehat{T}_2(u+2\eta) \cdot\ldots\cdot \widehat{T}_n(u+[n-1]\cdot 2\eta)\
                                                    P_{1\ldots n}^{\sss +}\notag\\[.5em]
   &=& \prod_{i=1}^N \frac{R_{(1\ldots n) q_i}(u,\eta)}{\zeta(u)\zeta(u+2\eta) \cdot\ldots\cdot \zeta(u+[n-1]\cdot 2\eta)}\ .
\end{eqnarray}
in the usual way by restriction to the only relevant matrixblock after applying the respective $A$-transformation. 
In the case of general open boundaries the fusion hierarchy for $n \geq 1$ is found to be
\begin{equation}
  \tau^\sss{(n)}(u)\cdot\tau^\sss{(1)}(u+n\cdot 2\eta) = -\frac{\tau^\sss{(n+1)}(u)}{\xi_{n}(u)}
          +\frac{\Delta(u+[n-1]\cdot 2\eta)}{\zeta(2u+2n\cdot 2\eta)}\cdot\xi_{n-1}(u)\tau^\sss{(n-1)}(u)
  \label{eq:OBCFusionHierarchyNoTilde}
\end{equation}
where $\Delta(u)$ labels the OBC super quantum determinant defined in (\ref{eqAP:OBCQuantumDeterminant}) and
\begin{equation}
  \xi_n(u) \equiv \prod_{k=1}^n \zeta(2u+[n+k]\cdot 2\eta)\ .
\end{equation}
The structure of this fusion hierarchy can be further simplified by introducing the rescaled quantities
\begin{equation}
  \tilde{\Delta}(u)          \equiv \frac{\Delta(u)}{\zeta(2u+2\cdot2\eta)} 
  \qquad\text{and}\qquad
  \tilde{\tau}^\sss{(n)}(u)  \equiv -\klammer{\prod_{i=1}^{n-1} \xi^{-1}_i(u)}\tau^\sss{(n)}(u)
  \label{eq:TauTilde}
\end{equation}
with the convenient definitions $\tilde{\tau}^\sss{(0)}(u) \equiv -\tau^\sss{(0)}(u) \equiv \eins$ and $\tilde{\tau}^\sss{(1)}(u) \equiv -\tau^\sss{(1)}(u)$ such that (\ref{eq:OBCFusionHierarchyNoTilde}) becomes
\begin{equation}
  \tilde{\tau}^\sss{(n)}(u)\cdot\tilde{\tau}^\sss{(1)}(u+n\cdot 2\eta)=\tilde{\tau}^\sss{(n+1)}(u)-\tilde{\Delta}(u+[n-1]\cdot 2\eta)\cdot\tilde{\tau}^\sss{(n-1)}(u)\ .
  \label{eq:OBCFusionHierarchyNice}
\end{equation}
\section{TQ-equations for OBC}

As in the PBC case, the fusion hierarchy (\ref{eq:OBCFusionHierarchyNice})
provides a system of relations between the eigenvalues
$\tilde{\Lambda}^{(n)}(u)$ of the fused (super) transfer matrices.  Defining
$\tilde{\Lambda}(u) \equiv \tilde{\Lambda}^{(1)}(u)$ and after shifting
$u\rightarrow u-n\cdot 2\eta$ this yields
\begin{equation}
	\tilde{\Lambda}(u) = \frac{\tilde{\Lambda}^{(n+1)}(u-n\cdot 2\eta)}{\tilde{\Lambda}^{(n)}(u-n\cdot 2\eta)}
	                     - \tilde{\Delta}(u-2\eta)\frac{\tilde{\Lambda}^{(n-1)}(u-n\cdot 2\eta)}{\tilde{\Lambda}^{(n)}(u-n\cdot 2\eta)}
\end{equation}
Introducing the functions
\begin{equation}
	h^{(n)}(u)\ \gamma^{(n)}(u)\ \tilde{Q}^{(n)}(u) \equiv \tilde{\Lambda}^{(n)}(u-n\cdot 2\eta) \\
\end{equation}
where
\begin{subequations}
\begin{eqnarray}
	\gamma^{(n)}(u) &\equiv& \frac{\sin(2u+2\eta)}{\sin(2u)} \prod_{j=1}^n \frac{\sin(2u-[2j-2]\cdot 2\eta)}{\sin(2u-[2j-3]\cdot 2\eta)} \\
	h^{(n)}(u)      &\equiv& -(-1)^n\prod_{k=0}^n \omega^\sss{+}\sin(u-k\cdot 2\eta - \psi_+)\cdot\omega^\sss{-}\sin(u-k\cdot 2\eta - \psi_-)
\end{eqnarray}
\end{subequations}
the eigenvalues can be written as
\begin{equation}
  \begin{split}
	\tilde{\Lambda}(u) =&\ \mathfrak{K}_{\delta}^\sss{+}(u) \mathfrak{K}_{\delta}^\sss{-}(u+2\eta) \frac{\sin(2u)}{\sin(2u+2\eta)}
	                     \frac{\tilde{Q}^{(n+1)}(u+2\eta)}{\tilde{Q}^{(n)}(u)} \\
	                     &-\frac{\tilde{\Delta}(u-2\eta)}{\mathfrak{K}_{\delta}^\sss{+}(u-2\eta) \mathfrak{K}_{\delta}^\sss{-}(u)}
	                     \frac{\sin(2u-2\eta)}{\sin(2u-4\eta)} \frac{\tilde{Q}^{(n-1)}(u-2\eta)}{\tilde{Q}^{(n)}(u)}
  \end{split}
\end{equation}
where the functions $\mathfrak{K}_{\alpha,\delta}^\sss{\pm}(u)$ are defined in
(\ref{eqAP:mathfrakKs}). Now assume that the limit $\tilde{Q}(u) \equiv
\lim_{n\rightarrow\infty} \tilde{Q}^{(n)}(u)$ exists and can be written as 
\begin{equation}
  \tilde{Q}(u) = f^N(u) \tilde{q}(u) \quad\text{with}\quad f(u)\equiv
  \mathrm{e}^{\mathrm{i}\pi \frac{u}{2\eta}}\frac{\sin(u-2\eta)}{\sin(u)}\ . 
\end{equation}
Resubstituting $\Lambda(u) = -\tilde{\Lambda}(u)$ by virtue of
(\ref{eq:TauTilde}) we obtain a TQ-equation for the open small polaron model
\begin{equation}
\label{eq:TQopen}
  \Lambda(u) = H_\alpha(u)\ \frac{\tilde{q}(u-2\eta)}{\tilde{q}(u)} -
  H_\delta(u)\ \frac{\tilde{q}(u+2\eta)}{\tilde{q}(u)}\ . 
\end{equation}
where the the functions $H_\alpha(u)$ and $H_\delta(u)$ factorize the super
quantum determinant (\ref{eqAP:OBCQuantumDeterminant}) as
\begin{equation}
\label{eq:Deltafact}
  H_\alpha(u) H_\delta(u-2\eta) = \zeta^{-1}(2u) \Delta(u-2\eta)\,.	
\end{equation}
As discussed in Appendix~\ref{app:SQD} the contribution of the boundary
matrices to the super quantum determinant $\Delta(u)$ of the small polaron
model is identical for diagonal and non-diagonal boundary fields.  Therefore,
$\Delta(u)$ can be factorized in the parametrization (\ref{eq:generalKs})
giving 
\begin{equation}
  \begin{split}
    H_\alpha(u) &\equiv \frac{\sin(2u+4\eta)}{\sin(2u+2\eta)}\
    \mathfrak{K}_{\alpha}^\sss{+}(u-2\eta) \mathfrak{K}_{\alpha}^\sss{-}(u) 
    \klammer{\frac{-\sin^2(u+2\eta)}{\sin(u+2\eta)\sin(u-2\eta)}}^N \\[.5em]
    H_\delta(u) &\equiv \frac{\sin(2u)}{\sin(2u+2\eta)}\
    \mathfrak{K}_{\delta}^\sss{+}(u) \mathfrak{K}_{\delta}^\sss{-}(u+2\eta) 
    \klammer{\frac{-\sin^2(u)}{\sin(u+2\eta)\sin(u-2\eta)}}^N \ .
  \end{split}
  \label{eq:ChosenQDetFactorization}
\end{equation}
With this factorization of the super quantum determinant the TQ-equation
(\ref{eq:TQopen}) coincides with the known result (\ref{eqAP:diagTQtype}) for
the diagonal boundary case obtained by means of the algebraic or coordinate
Bethe ansatz \cite{UmFW99,GuFY99,WaFG00}.  In this case the spectral problem
for the $M$-particle sector of the small polaron model can be solved using the
factorized ansatz (\ref{eqAP:diagQFunction})
\begin{equation}
\label{eq:Qfact}
  \tilde{q}(u)=\prod_{\ell=1}^M \sin(u+2\eta+\nu_\ell)\sin(u-\nu_\ell)
\end{equation}
where the unknown parameters $\nu_\ell$, $\ell=1,\ldots,M$ have to satisfy the
Bethe equations (\ref{eqAP:diagBethe}).



For generic non-diagonal boundary matrices an ansatz (\ref{eq:Qfact}) leads to
a constraint on the boundary parameters (and the number $M$) which guarantees
consistency between the asymptotic behaviour of the right hand side of
(\ref{eq:TQopen}) and the known behaviour of the transfer matrix eigenvalues
$\Lambda_\pm(u)$ (\ref{eq:NonDiagonalAsymptotics}).  Using such a requirement
Bethe equations have been formulated for the spectral problem of open
(non-diagonal) XXZ and XYZ Heisenberg spin chains \cite{YaNZ06,YaZh06,FrNR07}.
Unfortunately, in the present case of the small polaron model the
factorization (\ref{eq:ChosenQDetFactorization}) of the quantum determinant
does not reproduce the leading asymptotic behaviour of the transfer matrix
eigenvalues for any non-diagonal boundary fields.

To proceed with the solution of the TQ-equation (\ref{eq:TQopen}) one has to
find a different factorization of the quantum determinant satisfying
(\ref{eq:Deltafact}) or to modify the ansatz (\ref{eq:Qfact}) for the
$Q$-functions.
%
%
Based on the dependence of the transfer matrix on the off-diagonal boundary
parameters in various limits (\ref{eq:TauPropertiesSemiClassical}),
(\ref{eq:NonDiagonalAsymptotics}) and observations for small system sizes we
propose that the $Q$-functions can be written as
\begin{equation}
\label{eq:QAnsatz_OND}
  \tilde{q}(u) = q(u) + \rho(u)\cdot\klammer{\beta_+\alpha_- -
    \alpha_+\beta_-}\ 
\end{equation}
in the case of non-diagonal boundary conditions with $q(u)$ being the
factorized expression (\ref{eq:Qfact}) as in the diagonal case and another
unknown function $\rho(u)$ depending on the anisotropy $\eta$ and the diagonal
boundary parameters $\psi_\sss{\pm}$.
To determine $\rho(u)$ the ansatz (\ref{eq:QAnsatz_OND}) should be used in the
TQ-equation (\ref{eq:TQopen}) together with the analytical properties of the
transfer matrix eigenvalues, in particular their asymptotic behaviour
(\ref{eq:NonDiagonalAsymptotics}).

%
\section{Truncation of the OBC fusion hierarchy}
From here on, for the sake of readability, some of the functions introduced
above will be equipped with a second parameter indicating for them to be taken
at that particular value of the anisotropy $\eta$. For
instance, $K^\sss{\pm}(u,\rho) \equiv K^\sss{\pm}(u) |_{\eta\rightarrow \rho}$
and so on and so forth.

\subsection{$K$-matrix truncation}
It is convenient to define the following functions
\begin{eqnarray}
  \mu^\sss{\pm}_n(u) &\equiv& \pm \delta\gklammer{K^\sss{\pm}(\mp u-2\eta_n,\eta_n)}\frac{\sin(2\eta_n)}{\sin(2u-2\cdot 2\eta_n)}
  \prod_{k=2}^{2n}\frac{\sin(2u+k\cdot 2\eta_n)}{\sin(2\eta_n)}\\
  \nu^\sss{\pm}_n(u) &\equiv& \mp\frac{\omega_n^\sss{\pm}}{\mu^\sss{\pm}_n(u)} \klammer{\frac{\omega_n^\sss{\pm}}{2}}^n
  \sin([n+1][u\mp\psi_\sss{\pm}])
  \prod_{i=1}^n \prod_{j=1}^i \frac{\sin(2u+[i+j]\cdot 2\eta_n)}{\sin(2\eta_n)}\ .
\end{eqnarray}
where $\omega_n^{\pm} \equiv \omega^\sss{\pm}(\eta_n)$ and to introduce the
shorthand notations 
\begin{equation}
 \begin{split}
	\mathcal{K}^\sss{-}_{\ll n\gg}(u,\eta) &\equiv \sigma^z_{\ll n\gg}\cdot K^\sss{-}_{\ll 1\ldots n\gg}(u+2\eta) \\
	\mathcal{K}^\sss{+}_{\ll n\gg}(u,\eta) &\equiv K^\sss{+}_{\ll 1\ldots n\gg}(u+2\eta)\cdot \sigma^z_{\ll n\gg}\ .
  \end{split}
\end{equation}
The truncation identities for the boundary matrices can then be expressed as
\begin{equation}
  \begin{split}
	& C_{\ll 1\ldots n\gg}\ K^{\sss{\pm}}_{\ll 1\ldots n\gg}(u,\eta_{n-1}) \ C^{-1}_{\ll 1\ldots n\gg}\\
	& ~= \mu^{\sss{\pm}}_{n-1}(u)
	      \begin{pmatrix}
	         \nu^{\sss{\pm}}_{n-1}(\mp u) & ~ & ~ \\
	          ~ & B_{\ll 1\ldots n-2\gg} \mathcal{K}^\sss{\pm}_{\ll n-2 \gg}(u,\eta_{n-1}) B^{-1}_{\ll 1\ldots n-2\gg} & * \\
	          ~ & ~ & (\pm 1)^n\nu^{\sss{\pm}}_{n-1}(\pm u)
          \end{pmatrix}
 \end{split}
 \label{eq:CBoundaryMatrixTruncation}
\end{equation}

\subsection{OBC super transfer matrix truncation}
In order to be compatible with the truncation identities for the boundary matrices, the $R$-matrix truncation identities (\ref{eq:RMatrixTruncation}) need to be recast, this time employing the $C$ transformation matrices,
\begin{equation}
 \begin{split}
  & C_{\ll 1\ldots n \gg}\ R_{\ll 1\ldots n\gg q}(u,\eta_{n-1})\ C^{-1}_{\ll 1\ldots n \gg}\\
  & = \begin{pmatrix}
        -\mathcal{M}_{n-1}(u)\ \sigma^z_q & ~ & ~ \\
        ~  & \zeta(u)\ \sigma^z_{q}\ \mathcal{R}^{(n-2)}_q(u+2\eta_{n-1},\eta_{n-1}) & * \\
        ~  & ~ & \mathcal{M}_{n-1}(u)\ (\sigma^z_q)^{n-1}
      \end{pmatrix}
 \end{split}
 \label{eq:CRMatrixTruncation}
\end{equation}
where in slight contrast to definition (\ref{eq:RMatrixTruncationDefs})
\begin{equation}
  \mathcal{R}^{(n)}_q(u,\eta) \equiv B_{\ll 1\ldots n\gg}\ R_{\ll 1\ldots n\gg q}(u)\ B_{\ll 1\ldots n\gg}^{-1}
\end{equation}
such that for the single row monodromy matrix
\begin{eqnarray}
  \mathcal{T}^{(n)}(u,\eta) &\equiv& C_{\ll 1\ldots n \gg}\ 
                                        R_{\ll 1\ldots n\gg q_N}(u) \cdot \ldots \cdot
                                        R_{\ll 1\ldots n\gg q_2}(u) R_{\ll 1\ldots n\gg q_1}(u)\
                                     C^{-1}_{\ll 1\ldots n \gg}\\
   &\equiv& C_{\ll 1\ldots n \gg}\ T_{\ll 1\ldots n\gg}(u)\ C^{-1}_{\ll 1\ldots n \gg}
\end{eqnarray}
the truncation identity at $\eta = \eta_{n-1}$ reads
\begin{eqnarray}
  &&\hspace{-1em}\mathcal{T}^{(n)}(u,\eta_{n-1}) = \label{eq:CMonodromyMatrixTruncation}\\[.5em]
  &&\begin{pmatrix}
      [-\mathcal{M}_{n-1}(u)]^N \prod_{i=N}^1 \sigma^z_{q_i} & ~ & ~ \\[.5em]
      ~ & \zeta^N(u) \prod_{i=N}^1 \sigma^z_{q_i}\ \mathcal{R}^{(n-2)}_{q_i}(u+2\eta_{n-1},\eta_{n-1})  & * \\[.5em]
      ~ & ~ & [\mathcal{M}_{n-1}(u)]^N \prod_{i=N}^1 (\sigma^z_{q_i})^{n-1}
    \end{pmatrix}\notag
\end{eqnarray}
%
%
Again it is convenient to introduce the $C$-transformed object
\begin{equation}
  \widehat{\mathcal{T}}^{(n)}(u,\eta) \equiv C_{\ll 1\ldots n \gg}\ \widehat{T}_{\ll 1\ldots n\gg}(u)\ C^{-1}_{\ll 1\ldots n \gg} 
\end{equation}
to easily recognize the truncation identity
\begin{eqnarray}
  &&\hspace{-1em}\widehat{\mathcal{T}}^{(n)}(u+[n-1]\cdot 2\eta_{n-1},\eta_{n-1}) = \frac{1}{\zeta^N(u)\ \zeta^N(u+2\eta_{n-1})\ldots \zeta^N(u+[n-1]\cdot 2\eta_{n-1})}\times
  \label{eq:CHatMonodromyMatrixTruncation}\\[.5em]
  &&\times\begin{pmatrix}
      [-\mathcal{M}_{n-1}(u)]^N \prod_{i=1}^N \sigma^z_{q_i} & ~ & ~ \\[.5em]
      ~ & \zeta^N(u) \prod_{i=1}^N \sigma^z_{q_i}\ \mathcal{R}^{(n-2)}_{q_i}(u+2\eta_{n-1},\eta_{n-1})  & * \\[.5em]
      ~ & ~ & [\mathcal{M}_{n-1}(u)]^N \prod_{i=1}^N (\sigma^z_{q_i})^{n-1}
    \end{pmatrix}\ .\notag
\end{eqnarray}
Now that the individual truncation identities for all the objects involved in
the construction of the fused OBC super transfer matrix $\tau^{(n)}(u)$ are
known, it can be shown by simple matrix multiplication\footnote{Due to the
  cyclic invariance of the supertrace, all the matrix objects in
  (\ref{eq:obcFusedTransfermatrix}) may be conjugated by means of the
  $C$-transformation without changing the actual super transfer matrix.} of
(\ref{eq:CBoundaryMatrixTruncation}$^{\sss +}$),
(\ref{eq:CMonodromyMatrixTruncation}),
(\ref{eq:CBoundaryMatrixTruncation}$^{\sss -}$) and
(\ref{eq:CHatMonodromyMatrixTruncation}) that 
\begin{equation}
  \tau^{(n)}(u,\eta_{n-1}) = \sTr{\ll 1\ldots n\gg}{
                              \begin{pmatrix}
	                           X^{\sss +}  & ~ & ~ \\
	                                     ~ & Y & * \\
	                                     ~ & ~ & X^{\sss -}
	                          \end{pmatrix}
                             }
  = X^{\sss +} - \sTr{\ll 1\ldots n-2\gg}{Y} + (-1)^{n}\ X^{\sss -}
  \label{eq:obcTauTruncationFormal}
\end{equation}
with the placeholders $X^{\sss \pm}$ and $Y$ defined by
\begin{equation}
  X^{\sss \pm} \equiv (\pm 1)^n\ \eckklammer{\prod_{k=0}^{n-1} \zeta^{-N}(u+k\cdot 2\eta_{n-1}) } \mathcal{M}_{n-1}^{2N}(u)\
                      \mu^{\sss +}_{n-1}(u)\mu^{\sss -}_{n-1}(u)\ \nu^{\sss +}_{n-1}(\mp u)\nu^{\sss -}_{n-1}(\pm u)
\end{equation}
and
\begin{equation}
  \begin{split}
   Y =&\ \phi^{\tau}_{n-1}(u)\
              B_{\ll 1\ldots n-2\gg}\  K^{\sss +}_{\ll 1\ldots n-2\gg}(u+2\eta_{n-1})\ \sigma^z_{\ll n-2 \gg} \klammer{\prod_{i=1}^N \sigma^z_{q_i}} \times \\
                          ~ & \times   T_{\ll 1\ldots n-2\gg}(u+2\eta_{n-1})\ \sigma^z_{\ll n-2 \gg}\
                                      K^{\sss -}_{\ll 1\ldots n-2\gg}(u+2\eta_{n-1}) \times \\
                          ~ & \times  \klammer{\prod_{i=1}^N \sigma^z_{q_i}} \widehat{T}_{\ll 1\ldots n-2\gg}(u+2\eta_{n-1}+[(n-2)-1]\cdot 2\eta_{n-1})\
              B^{-1}_{\ll 1\ldots n-2\gg}\\[.5em]
	 =&\ \phi^{\tau}_{n-1}(u)\
	           B_{\ll 1\ldots n-2\gg}\ K^{\sss +}_{\ll 1\ldots n-2\gg}(u+2\eta_{n-1})\  T_{\ll 1\ldots n-2\gg}(u+2\eta_{n-1}) \times \\
	                      ~ & \times K^{\sss -}_{\ll 1\ldots n-2\gg}(u+2\eta_{n-1})\ \widehat{T}_{\ll 1\ldots n-2\gg}(u+2\eta_{n-1}+[(n-2)-1]\cdot 2\eta_{n-1})\
	          B^{-1}_{\ll 1\ldots n-2\gg}\ .
  \end{split}
  \label{eq:Ydefinition}
\end{equation}
%
%
In the second step of equation (\ref{eq:Ydefinition}) relation
(\ref{eqAP:fusedQDetCommutation}) has been employed to get rid of the
$\sigma^z$ factors such that (\ref{eq:obcTauTruncationFormal}) eventually
yields the truncation identities for the OBC transfer matrices, 
\begin{equation}
 \begin{split}
	\tau^{(n)}(u,\eta_{n-1}) = & \phi^{\text{id}}_{n-1}(u) \cdot \eins - \phi^{\tau}_{n-1}(u) \cdot \tau^{(n-2)}(u+2\eta_{n-1},\eta_{n-1})
 \end{split}
 \label{eq:obcTauTruncation}
\end{equation}
where $\phi^{\text{id}}_{n}(u)$ and $\phi^{\tau}_{n}(u)$ are rather lengthy
expressions given by 
\begin{equation}
  \begin{aligned}
    \phi^{\text{id}}_{n}(u)   =& \eckklammer{\prod_{k=0}^{n}
      \zeta^{-N}(u+k\cdot 2\eta_{n}) } \mathcal{M}_{n}^{2N}(u)\ \mu^{\sss
      +}_{n}(u)\mu^{\sss -}_{n}(u)\ 
    [ \nu^{\sss +}_{n}(-u)\nu^{\sss
      -}_{n}(u) + \nu^{\sss
      +}_{n}(u)\nu^{\sss -}_{n}(-u) ] \\ 
    \phi^{\tau}_{n}(u) =& \klammer{\frac{\zeta(u)}{\zeta(u+n\cdot
        2\eta_{n})}}^N \mu^{\sss +}_{n}(u)\mu^{\sss -}_{n}(u)\ . 
  \end{aligned}
\end{equation}
In terms of the rescaled transfer matrices (\ref{eq:OBCFusionHierarchyNice})
it is reasonable to introduce 
\begin{equation}
 \begin{aligned}
   \tilde\phi^{\text{id}}_{n}(u)  =& -\eckklammer{\prod_{i=1}^{n}
     \xi^{-1}_i(u)}_{\eta=\eta_n} \phi^{\text{id}}_{n}(u)\\ 
   \tilde\phi^{\tau}_{n}(u)       =& \eckklammer{\prod_{i=1}^{n}
     \xi^{-1}_i(u)}_{\eta=\eta_n} \phi^{\tau}_{n}(u)
   \eckklammer{\prod_{i=1}^{n-2} \xi_i(u+2\eta_n)}_{\eta=\eta_n}\ . 
\end{aligned}
\end{equation}
which yield the respective rescaled truncation identities
\begin{equation}
 \begin{split}
	\tilde\tau^{(n)}(u,\eta_{n-1}) = & \tilde\phi^{\text{id}}_{n-1}(u) \cdot \eins - \tilde\phi^{\tau}_{n-1}(u) \cdot \tilde\tau^{(n-2)}(u+2\eta_{n-1},\eta_{n-1})\ .
 \end{split}
 \label{eq:obcTauTruncationRescaled}
\end{equation}
%
%
\section{Summary and Conclusion}
Starting from structures provided by the Yang-Baxter algebra (\ref{eq:YBA})
and the reflection algebra (\ref{eq:RA1}), (\ref{eq:RA2}) we have set up the
fusion hierarchies for the commuting transfer matrices $\tau^{(n)}(u)$ of the
small polaron model with periodic and general open boundary conditions,
respectively.  Following previous work on spin chains with non-diagonal
boundary fields \cite{YaNZ06,YaZh06,FrNR07} we have obtained TQ-equations for
the eigenvalues of the transfer matrices by assuming the limit $n\to\infty$ of
these expressions to exist.  These TQ-equations can be solved by functional
Bethe ansatz methods in the case of periodic and diagonal open boundary
conditions.  The resulting spectrum coincides with what has been found
previously using the algebraic Bethe ansatz
\cite{PuZh86,ZhJW89,UmFW99,GuFY99,WaFG00} and was to be expected as a
consequence of the Jordan-Wigner equivalence of the small polaron model with
the spin-1/2 XXZ Heisenberg chain.

For generic non-diagonal boundary conditions the $U(1)$ symmetry of the model
corresponding to particle number conservation is broken.  Therefore, the
algebraic approach cannot be applied as it uses the Fock vacuum as a reference
state and relies on this being an eigenstate of the system.  This situation is
well known from the (ungraded) spin-1/2 XXZ Heisenberg chain with non-diagonal
boundary fields where in spite of significant activities a practical solution
of the eigenvalue problem for generic anisotropies and boundary fields is
lacking.  Here we have used the strategies employed previously for the XXZ
chain to the (graded) small polaron chain: 
apart from the formulation of the spectral problem in terms of a TQ-equation
the fusion hierarchy can be truncated at a finite order for anisotropies
being roots of unity, $\eta_p=\pi/(2(p+1))$ \cite{Nepo02}.  We have derived the
corresponding truncation identities for the small polaron model subject to all
boundary conditions considered.
Inspection of the $R$-matrices obtained at the first few fusion levels
suggests that it is possible to derive similar identities for anisotropies
given by integer multiples of $\eta_p$.

To actually compute eigenvalues of the transfer matrices further steps have to
be taken: for anisotropies being roots of unity the truncated fusion hierarchy
can be be analyzed following the steps which have been established for the XXZ
chain \cite{MuNe05,MuNS06,MuNS07} where additional constraints on the boundary
fields may arise.  For generic anisotropies the situation is more complicated:
in the ungraded XXZ chain a (factorized) Bethe ansatz for the $Q$-function
given in terms of finitely many parameters such as (\ref{eq:Qfact}) was
possible only if the boundary parameters satisfy a constraint
\cite{CaoX03,Nepo04,YaNZ06,FrNR07,FrSW08}.
For graded models such a constraint may be absent: in the rational limit
$\eta\to0$ of the model considered here the functional form of the
$Q$-function remained unchanged when off-diagonal boundary fields where added
\cite{GrFr10}.  
Similarly, the nilpotency of the off-diagonal boundary fields may allow for a
general solution of the small polaron model.  As shown in
Appendix~\ref{app:SQD}, the super quantum determinant of this model depends
only on the diagonal boundary parameters which simplifies the factorization
problem (\ref{eq:Deltafact}).  In addition, the odd Grassmann numbers
parametrizing the off-diagonal boundary fields appear only in a specific
combination.
Therefore, starting with the proposed ansatz (\ref{eq:QAnsatz_OND}) for the
$Q$-function the derivation of Bethe type equations appears to be possible in
the generic case.  These open questions shall be addressed in a future
publication.

A possible extension of the present work is to consider integrable higher spin
chains with generic boundary conditions.  Such generalizations of an
integrable model can be constructed by application of the fusion method
\cite{KuRS81,MeNe92a,Zhou96b} in the quantum spaces of the model in addition
to fusion in auxiliary space as used in this paper for the derivation of the
fusion hierarchies (\ref{eq:FusionHierarchyPBC}) and
(\ref{eq:OBCFusionHierarchyNice}).  Starting from the spin-1/2 XXZ Heisenberg
chain this leads to the hierarchy of integrable higher spin XXZ models
\cite{KuRe83a,KiRe86,KiRe87} including the spin-1 Fateev-Zamolodchikov model
\cite{ZaFa80,MeNR90}.  Similarly, this method has been used for the
construction and solution of graded models based on higher spin
representations of super Lie algebras, see e.g.\
\cite{Maas95,PfFr96,PfFr97,Frahm99,Grun00}.  In the present context this would
lead to integrable generalizations of the small polaron model with general
boundary conditions.
The local Hilbert spaces of these models have dimension $(n/2|n/2)$ for $n$
even and $((n+1)/2|(n-1)/2)$ for $n$ odd, see Table~\ref{tb:AuxSpaceGrading}.
A quantum chain with local interactions can be constructed from $R$-matrices
acting on the tensor product of two copies of such a space.  The integrable
open boundary conditions for these models are given in terms of the fused
$K$-matrices (\ref{eq:kmatm}).  Taking into account the gradation the possible
states can be identified e.g.\ with the internal degrees of freedom of a
fermionic lattice model with several local orbitals to allow for a physical
interpretation of the resulting quantum chain.  The higher spin XXZ models
with general open boundary conditions the spectral problem has been studied by
Frappat \emph{et al.}  \cite{FrNR07} who found that the solution requires
similar constraints as in the spin-1/2 case.

%
\begin{acknowledgments}
  We like to thank Nikos Karaiskos for useful discussions on the subject of
  this paper.  This work has been supported by the Deutsche
  Forschungsgemeinschaft under grant no.\ Fr~737/6.
\end{acknowledgments}

\newpage
\appendix
\section{Graded vector spaces}
\label{app:GradedVectorSpaces}
Fermionic lattice models exhibit a natural $\mathbb{Z}_2$ gradation on their
local space of states, i.e. $V=V_0\oplus V_1$ is equipped with a notion of
parity,
\begin{equation}
  \operatorname{p}: V_i \rightarrow \mathbb{Z}_2 \quad,\quad
  \operatorname{p}(v_i) \mapsto i\in\{0,1\}\ . 
\end{equation}
Let $\dim V_0 \equiv m\in\mathbb{N}$ and $\dim V_1 \equiv n\in\mathbb{N}$ be
finite. Then $V$ is said to have dimension $(m|n)$ and $V_0$, $V_1$ are called
the \herv{homogeneous subspaces} of $V$. An element $v\in V$ is said to be
\herv{even} if $\parity(v)=0$ and is respectively called \herv{odd} if
$\operatorname{p}(v)=1$. While even elements of $V$ correspond to bosonic
states, odd elements represent fermionic states.
For instance consider the case where both of the homogeneous subspaces $V_0$
and $V_1$ are one-dimensional such that the composite local space of states
$V=V_0\oplus V_1$ is spanned by just one bosonic and one fermionic state. Then
$V$ is said to have $BF$-grading, where $BF$ refers to an ordered basis of $V$
in which the first basis vector is associated to with the bosonic state ($B$)
whereas the second basis vector is associated to the fermionic state ($F$).
Now consider the tensor product of two copies of $V$. Taking into account the
order of the basis states, the tensor product space will have $BFFB$-grading,
\begin{equation}
  V\otimes V = (V_0\oplus V_1)\otimes(V_0\oplus V_1)
  = \underbrace{(V_0\otimes
    V_0)}_{B}\oplus\underbrace{(V_0\otimes
    V_1)}_{F}\oplus\underbrace{(V_1\otimes
    V_0)}_{F}\oplus\underbrace{(V_1\otimes V_1)}_{B}\ . 
\end{equation}
In the following, the conventions from \cite{GoMu98} will essentially be
adopted. 

Let $\{ e_1, e_2,\ldots, e_m, e_{m+1},\ldots, e_{m+n} \}$ be a homogeneous
basis of $V$, i.e. each basis element has distinct parity $\parity(e_\alpha)$,
and for convenience let this basis be ordered, such that the first $m$
elements span the even and the last $n$ elements span the odd subspace of $V$, 
\begin{equation}
  \parity(\alpha)\equiv\parity(e_\alpha)=\begin{cases}
    0 & \text{if}\quad 1\leq\alpha\leq m
    \\ 
    1 & \text{if}\quad m+1\leq\alpha\leq m+n
  \end{cases}\ .
\end{equation}
In order to deal with an algebra of linear operators, acting on the graded
local space of states, it is necessary to extend the concept of parity to
$\End(V)$, the space of endomorphisms of $V$. The $(m+n)\times(m+n)$ basis
elements of $\End(V)$ will be labeled $e_\alpha^{~\beta}$ and are defined
through their action on the above basis of $V$, 
\begin{equation}
	e_\alpha^{~\beta} e_\gamma \equiv \delta^\beta_\gamma e_\alpha\ .
\end{equation}
By extending the definition of the parity function to
\begin{equation}
	\parity(e_\alpha^{~\beta}) \equiv \parity(\alpha) + \parity(\beta) \mod 2
\end{equation}
$\End(V)$ becomes a $\mathbb{Z}_2$ graded vector space. A basis of the $N$-fold product space
\begin{equation}
 \End^{\otimes N}(V) \equiv \underbrace{\End(V)\otimes\End(V)\otimes\ldots\otimes\End(V)}_{N \text{ times}}
\end{equation}
can most naturally be obtained by embedding the local basis elements
$e_\alpha^{~\beta}$ into this tensorproduct structure. Moreover,
$\End^{\otimes N}(V)$ aquires a $\mathbb{Z}_2$ grading by a further extension
of the definition of the parity function,
\begin{equation}
	\parity(e_{\alpha_1}^{~\beta_1}\otimes e_{\alpha_2}^{~\beta_2}\otimes\ldots\otimes e_{\alpha_N}^{~\beta_N})
	\equiv \parity(e_{\alpha_1}^{~\beta_1}) + \parity(e_{\alpha_2}^{~\beta_2}) + \ldots + \parity(e_{\alpha_N}^{~\beta_N}) \mod 2\ .
\end{equation}
When dealing with graded vector spaces, it is useful to replace the usual
tensorproduct structure by a so-called \herv{super tensorproduct}. The symbol
$\sTP$ will be used to distinguish this new structure. With respect to a
certain basis, the components of the super tensorproruct of two operators
$A\in\End^{\otimes k}(V)$ and $B\in\End^{\otimes l}(V)$, where
$k,l\in\mathbb{N}$, is explicitly defined through 
\begin{equation}
	(A\sTP B)^{\alpha\gamma}_{~\beta\delta} = (-1)^{[\parity(\alpha)+\parity(\beta)]\parity(\gamma)}A^\alpha_{~\beta} B^\gamma_{~\delta}\ .
\end{equation}
As pointed out in \cite{GoMu98}, the super tensorproduct allows for a most
convenient \herv{graded} embedding of the $e_\alpha^{~\beta}$ into the $j$-th
subspace of $\End^{\otimes N}(V)$, 
\begin{eqnarray}
	e_j,_\alpha^{\beta} \equiv \eins^{\sTP(j-1)}\sTP e_\alpha^{~\beta}\sTP \eins^{\sTP(N-j)}\ .
	\label{eqAP:sTPembeddedBasis}
\end{eqnarray}
%
A graded version of the permutation operator $\mathcal{P}$ is defined by the relation
\begin{equation}
	\mathcal{P}(A\sTP B) = (B\sTP A)\mathcal{P}\ .
\end{equation}
If (\ref{eqAP:sTPembeddedBasis}) is employed as a basis for $\End^{\otimes N}(V)$, the operator $\mathcal{P}_{ij}$ which permutes the $i$-th and the $j$-th subspace can explicitly be constructed as
\begin{equation}
	\mathcal{P}_{ij} = (-1)^{\parity(\beta)} e_i,_\alpha^{~\beta} e_j,_\beta^{~\alpha}\ .
\end{equation}
%
In the following, the definitions of some well-known operations, namely the
matrix transposition and the trace operation, will be adapted to fit the needs
of graded vector spaces. A nicely motivated and much more elaborated list of
matrix operations on graded vector spaces can found in \cite{CORNWELL3}.  
\begin{itemize}
\item Firstly, the \herv{super transposition} of an element $A\in\End(V)$ is
  defined by 
  \begin{equation}
    (A^\st)^\alpha_{~\beta} = (-1)^{\parity(\alpha)[\parity(\alpha+\beta)]} A_\beta^{~\alpha}\ .
  \end{equation}
  In contrast to the ungraded case, the super transposition is not an
  involution, i.e. applying the super transposition twice does not yield the
  identity operation.  As pointed out in \cite{BGZZ98}, it is therefore
  convenient to introduce an \herv{inverse super transposition},
  \begin{equation}
    (A^\ist)^\alpha_{~\beta} = (-1)^{\parity(\beta)[\parity(\alpha+\beta)]} A_\beta^{~\alpha}\ .
  \end{equation}
  The \herv{partial} super transposition, i.e. a super transposition on the
  $j$-th subspace of $\End^{\otimes N}(V)$, is defined through
  \begin{equation}
    (A_1 \sTP \ldots \sTP A_j \sTP \ldots \sTP A_N)^{\st_j} \equiv A_1 \sTP
    \ldots \sTP (A_j)^\st \sTP \ldots \sTP A_N \ . 
  \end{equation}
  The partial inverse super transposition is defined analogously. Please note
  that, as opposed to ordinary partial matrix transpositions on ungraded
  vector spaces, the successive application of partial super transpositions on
  all subspaces is gernerally not equal to a total super transposition,
  i.e. $(A_1 \sTP A_2)^{\st_1 \st_2} \neq (A_1\sTP A_2)^\st$ .
\item Secondly, the \herv{super trace} of some $A\in\End(V)$ is given by
  \begin{equation}
    \str{A} \equiv \sum_\alpha(-1)^{\parity(\alpha)} A^{\alpha}_{~\alpha}\ .
  \end{equation}
  For operators $B\in\End^{\otimes N}(V)$ it is convenient to define a
  \herv{partial super trace} on subspace $j$ as 
  \begin{equation}
    \sTr{j}{B}^{\alpha_1\, \ldots \alpha_{j-1}\, \alpha_{j+1}\, \ldots \,
      \alpha_N}_{~\beta_1\, \ldots \beta_{j-1}\, \beta_{j+1}\, \ldots \,
      \beta_N} 
    \equiv \sum_\gamma(-1)^{\parity(\gamma)}
    B^{\alpha_1\, \ldots \alpha_{j-1}\, \gamma\; \alpha_{j+1}\, \ldots \,
      \alpha_N}_{~\beta_1\, \ldots \beta_{j-1}\, \gamma\; \beta_{j+1}\, \ldots
      \, \beta_N}\ . 
  \end{equation}
\end{itemize}
%

\section{Relation to Bracken's dual reflection algebra}
\label{app:RelationToBracken}
According to \cite{BGZZ98} the dual reflection equation for quite general graded models reads
\begin{equation}
 \begin{split}
	R_{12}(v-u) K_1^\sss{+}(u) &\widetilde{\widetilde{R}}_{21}(-u-v)^{\ist_1 \st_2} K_2^\sss{+}(v)  \\[.5em]
	=\ &K_2^\sss{+}(v) \widetilde{R}_{12}(-u-v)^{\ist_1 \st_2} K_1^\sss{+}(u) R_{21}(v-u)
 \end{split}
 \label{eqAP:gdRE1}
\end{equation}
where
\begin{eqnarray}
	\widetilde{\widetilde{R}}_{21}(\lambda)^{\ist_1 \st_2} &=& \klammer{\eckklammer{\gklammer{R^{-1}_{21}(\lambda)}^{\ist_2}}^{-1}}^{\st_2}\\[.5em]
	\widetilde{R}_{12}(\lambda)^{\ist_1 \st_2}             &=& \klammer{\eckklammer{\gklammer{R^{-1}_{12}(\lambda)}^{\st_1}}^{-1}}^{\ist_1}\ .
\end{eqnarray}
By performing a super transposition on the first space and an inverse super transposition on the second, i.e. by applying $(.)^{\st_1\ist_2}$ to equation (\ref{eqAP:gdRE1}) one obtains the equivalent form
\begin{equation}
 \begin{split}
	R_{21}(v-u)^{\st_1\ist_2} K_1^\sss{+}(u)^{\st_1} &\widetilde{R}_{12}(-u-v) K_2^\sss{+}(v)^{\ist_2}  \\[.5em]
	=\ &K_2^\sss{+}(v)^{\ist_2} \widetilde{\widetilde{R}}_{21}(-u-v) K_1^\sss{+}(u)^{\st_1} R_{12}(v-u)^{\st_1\ist_2}
 \end{split}
 \label{eqAP:gdRE2}
\end{equation}
In the case of the small polaron R-matrix as defined in (\ref{eq:RMatrix}) one finds
 \begin{eqnarray}
	\widetilde{\widetilde{R}}_{21}(\lambda) &=& \frac{\zeta(\lambda)}{\zeta(\lambda-2\eta)}\ R_{12}(\lambda-4\eta)\\[.5em]
	\widetilde{R}_{12}(\lambda)             &=& \frac{\zeta(\lambda)}{\zeta(\lambda-2\eta)}\ R_{21}(\lambda-4\eta)
\end{eqnarray}
At this point it is convenient to introduce a shorthand, which will henceforth be referred to as \herv{conjugated R-matrix},
\begin{equation}
    \bar{R}_{ba}(\lambda) \equiv M_a^{-1}\ R_{ba}(\lambda)\ M_a
    \label{eqAP:MConjugatedR}
\end{equation}
with $M$ being the so-called crossing matrix. For the small polaron model in particular, it is found that $M=\sigma^z$ such that
\begin{equation}
	\begin{split}
   	 \bar{R}_{ab}(\lambda) &= R_{ba}^{\st_a\ist_b}(\lambda) \stackrel{\sss{(\ref{eq:RMatrixPropertiesPSymmetry})}}{=} R_{ba}^{\ist_a\st_b}(\lambda) \\[.5em]
                                &= R_{ab}^{\st^2_a}(\lambda) = R_{ab}^{\ist^2_a}(\lambda) = R_{ab}^{\st^2_b}(\lambda) = R_{ab}^{\ist^2_b}(\lambda)\ .
    \end{split}
    \label{eqAP:ConjugatedR}
\end{equation}
Using this conjugated R-matrix (\ref{eqAP:ConjugatedR}), the dual reflection equation may be written as
\begin{equation}
 \begin{split}
	\bar{R}_{12}(v-u) K_1^\sss{+}(u)^{\st_1} & R_{21}(-u-v-4 \eta) K_2^\sss{+}(v)^{\ist_2}  \\[.5em]
	=\ &K_2^\sss{+}(v)^{\ist_2} R_{12}(-u-v-4 \eta) K_1^\sss{+}(u)^{\st_1} \bar{R}_{21}(v-u)
 \end{split}
 \label{eqAP:gdRE3}
\end{equation}
and is graphically depicted by \bigskip
\begin{center}
  \includegraphics[width=0.75\textwidth]{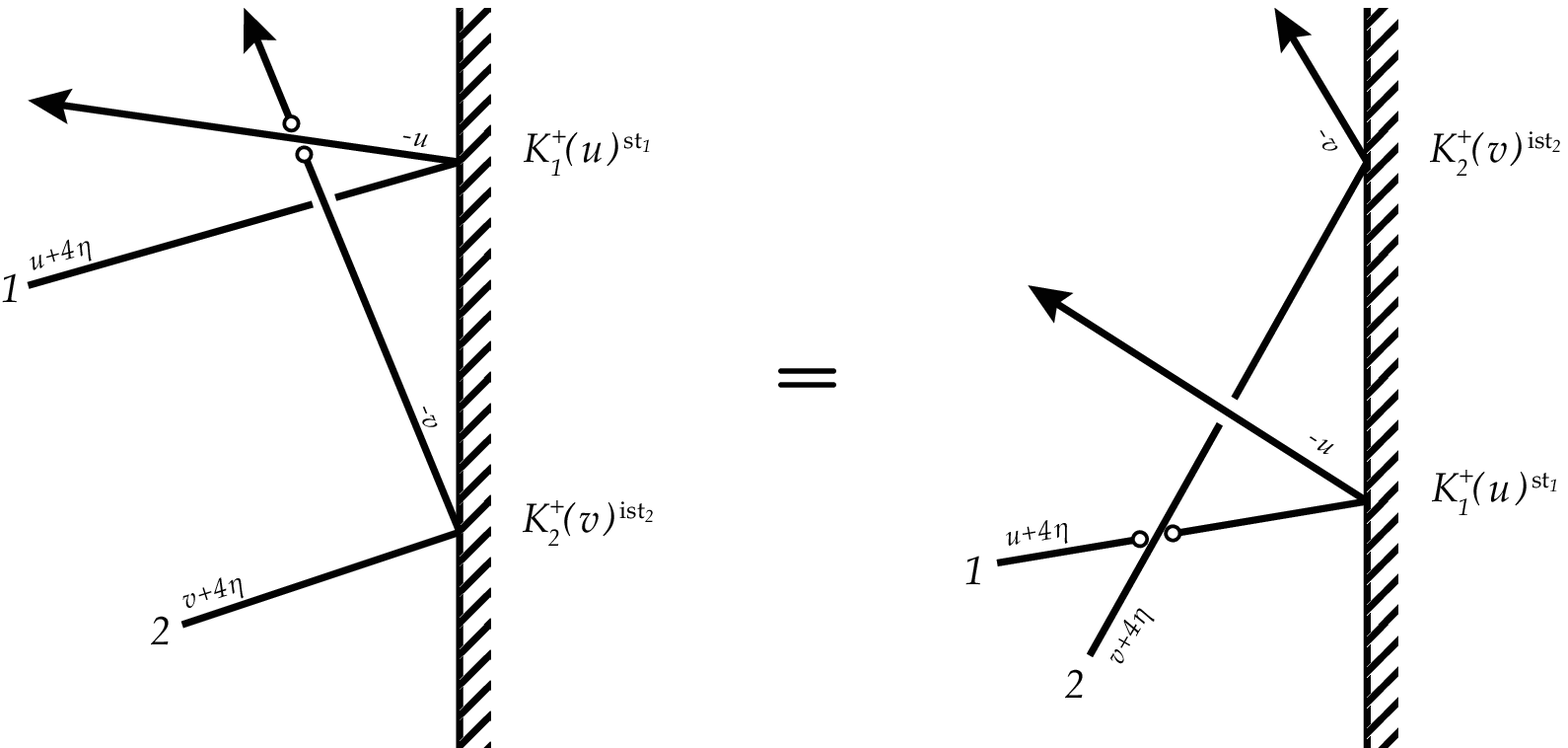}
\end{center}

%
\section{Algebraic Bethe ansatz for diagonal boundaries}
\label{app:diagABA}
The reflection equation (\ref{eq:RA1}) gives 16 fundamental commutation relations for the quantum space operators $\calA$, $\calB$, $\calC$ and $\calD$ of which the following three are of particular interest,
\begin{eqnarray}
&&\begin{aligned}
 \calB(u)\calB(v)=&\calB(v)\calB(u)
\end{aligned}\\[1em]
&&\begin{aligned}
 \calA(u)\calB(v)=&\textstyle\frac{\sine{0}{u+v}\sine{2}{v-u}}{\sine{0}{v-u}\sine{2}{u+v}}\displaystyle\calB(v)\calA(u)\\
 &~+\textstyle\frac{\vartheta(v)\sine{2}{0}}{\sine{2}{u+v}}\displaystyle\calB(u)\gklammer{\textstyle\frac{\sine{0}{2v}\sine{2}{u+v}}{\vartheta(v)\sine{0}{u-v}\sine{2}{2v}}\displaystyle\calA(v)-\widetilde{\calD}(v)}
\end{aligned}\\[1em]
&&\begin{aligned}
 \widetilde{\calD}(u)\calB(v)=&\textstyle\frac{\sine{4}{u+v}\sine{2}{u-v}}{\sine{0}{u-v}\sine{2}{u+v}}\displaystyle\calB(v)\widetilde{\calD}(u)-\textstyle\frac{\sine{2}{0}\sine{4}{2u}\sine{0}{2v}}{\vartheta(u)\sine{2}{2u}\sine{2}{u+v}\sine{2}{2v}}\displaystyle\calB(u)\times\qquad\qquad\\
 &~\times\gklammer{\textstyle\frac{\vartheta(v)\sine{2}{2v}\sine{2}{u+v}}{\sine{0}{2v}\sine{0}{u-v}}\displaystyle\widetilde{\calD}(v)-\calA(v)}\, ,
\end{aligned}
\end{eqnarray}
using the abbreviation $\sine{k}{\lambda}\equiv\sin(\lambda+k\eta)$. To obtain
the desired commutation relations, it is necessary to make an ansatz for a
shifted $\calD$-operator
\begin{equation}
  \calD(\lambda)=\vartheta(\lambda)\widetilde{\calD}(\lambda)
  +\phi(\lambda)\calA(\lambda)  
\end{equation}
and to determine the scalar functions $\phi(\lambda)$ and $\vartheta(\lambda)$. It turns out that $\phi(\lambda)=\frac{\sine{2}{0}}{\sine{2}{2\lambda}}$ while $\vartheta(\lambda)$ remains arbitrary.
Starting from the general boundary matrices given in (\ref{eq:generalKs}) the
diagonal case can easily be obtained by setting
$\alpha_\sss{\pm}=\beta_\sss{\pm}=0$. This leads to Bethe equations 
\begin{equation}
\label{eqAP:diagBethe}
 \klammer{\frac{\sine{2}{\nu_j}}{\sine{0}{\nu_j}}}^{2N}=\frac{\sine{2}{\nu_j-\psi_\sss{+}}\sine{2}{\nu_j-\psi_\sss{-}}}{\sine{0}{\nu_j+\psi_\sss{+}}\sine{0}{\nu_j+\psi_\sss{-}}}
 \prod_{\stackrel{\scriptstyle\ell=1}{\scriptstyle\ell\neq j}}^M\frac{\sine{4}{\nu_j+\nu_\ell}\sine{2}{\nu_j-\nu_\ell}}{\sine{0}{\nu_j+\nu_\ell}\sine{-2}{\nu_j-\nu_\ell}}
\end{equation}
and super transfer matrix eigenvalues\footnote{Note, that this result
  corresponds to the one obtained by \cite{UmFW99}. However, the authors of
  \cite{UmFW99} seem to have made a slight mistake when substituting their
  formula (57) into (61) to obtain (62), which should correctly
  read $$t(u)=\boldsymbol{+}\frac{\sin(2u+4\eta)\sin(u\boldsymbol{+}t^+)}{\sin(2u+2\eta)}\calA(u)\boldsymbol{-}\frac{\sin(u+2\eta\boldsymbol{-}t^+)}{\sin(2u+2\eta)}\tilde{\calD}(u)\,. $$
}
\begin{equation}
 \begin{split}
 &\Lambda(u)=\mathfrak{K}_\alpha^\sss{-}(u)\klammer{\mathfrak{K}_\alpha^\sss{+}(u)-\frac{\sine{2}{0}}{\sine{2}{2u}}\mathfrak{K}_\delta^\sss{+}(u)}\klammer{\frac{\sine{2}{u}}{\sine{2}{-u}}}^N
 \prod_{\ell=1}^M \frac{\sine{0}{u+\nu_\ell}\sine{2}{\nu_\ell-u}}{\sine{0}{\nu_\ell-u}\sine{2}{u+\nu_\ell}}\\
 &-\mathfrak{K}_\delta^\sss{+}(u)\klammer{\mathfrak{K}_\delta^\sss{-}(u)-\frac{\sine{2}{0}}{\sine{2}{2u}}\mathfrak{K}_\alpha^\sss{-}(u)}\klammer{\frac{\sine[2]{0}{u}}{\sine{2}{u}\sine{2}{-u}}}^N \prod_{\ell=1}^M \frac{\sine{4}{u+\nu_\ell}\sine{2}{u-\nu_\ell}}{\sine{0}{u-\nu_\ell}\sine{2}{u+\nu_\ell}}\, .\quad
 \end{split}
 \label{eqAP:diagEigenvalue}
\end{equation}
Here $\mathfrak{K}_{\alpha,\delta}^\sss{\pm}(u)$ label the diagonal entries of the boundary matrices (\ref{eq:generalKs}),
\begin{equation}
 \begin{array}{lclr}
  \mathfrak{K}_{\alpha}^\sss{-}(u) = \omega^\sss{-}\sin(\psi_\sss{-}+u) &,& \mathfrak{K}_{\alpha}^\sss{+}(u) = \omega^\sss{+}\sin(u+2\eta+\psi_\sss{+})&,\\[1em]
  \mathfrak{K}_{\delta}^\sss{-}(u) = \omega^\sss{-}\sin(\psi_\sss{-}-u )&,& \mathfrak{K}_{\delta}^\sss{+}(u) = \omega^\sss{+}\sin(u+2\eta-\psi_\sss{+})&.
 \end{array}
 \label{eqAP:mathfrakKs}
\end{equation}
Introducing the functions
\begin{equation}
 q(u) \equiv \prod_{\ell=1}^M \sin(u+2\eta+\nu_\ell)\sin(u-\nu_\ell)
 \label{eqAP:diagQFunction}
\end{equation}
the eigenvalues (\ref{eqAP:diagEigenvalue}) can be recast as
\begin{equation}
 \begin{split}
 \Lambda(u)q(u)&=\mathfrak{K}_\alpha^\sss{-}(u)\klammer{\mathfrak{K}_\alpha^\sss{+}(u)-\frac{\sine{2}{0}}{\sine{2}{2u}}\mathfrak{K}_\delta^\sss{+}(u)}\klammer{\frac{\sine[2]{2}{u}}{\sine{2}{u}\sine{2}{-u}}}^N
 q(u-2\eta)\\
 &-\mathfrak{K}_\delta^\sss{+}(u)\klammer{\mathfrak{K}_\delta^\sss{-}(u)-\frac{\sine{2}{0}}{\sine{2}{2u}}\mathfrak{K}_\alpha^\sss{-}(u)}\klammer{\frac{\sine[2]{0}{u}}{\sine{2}{u}\sine{2}{-u}}}^N q(u+2\eta)\, .\quad
 \end{split}
 \label{eqAP:diagTQtype}
\end{equation}
%
%

\section{Super quantum determinants}
\label{app:SQD}
Consider a generic $BFFB$ graded $R$-matrix of the shape
\begin{equation}
	R(u) = \begin{pmatrix}
	        a(u+2\eta) & 0        & 0        & 0          \\
	        0          & a(u)     & a(2\eta) & 0          \\
	        0          & a(2\eta) & a(u)     & 0          \\
	        0          & 0        & 0        & -a(u+2\eta)
           \end{pmatrix}
\end{equation}
where $a(-u) = a(u)$ and $a(0)=0$. At $u=-2\eta$ such an $R$-matrix gives rise to a projector $P^\sss{-}$ onto a one-dimensional subspace
\begin{equation}
    P^\sss{-} = -\frac{1}{2 a(2\eta)} R(-2\eta)
              = \begin{pmatrix}
	              0 & 0    & 0    & 0 \\
	              0 & 1/2  & -1/2 & 0 \\
	              0 & -1/2 & 1/2  & 0 \\
	              0 & 0    & 0    & 0
                \end{pmatrix}\ .
\end{equation}
Let $T(u)$ be a representation of the graded YBA
\begin{equation}
	R_{12}(u-v)\ T_1(u)\ T_2(v) = T_2(v)\ T_1(u)\ R_{12}(u-v)
\end{equation}
with the usual embeddings $T_1(u)\equiv T(u)\sTP \eins$ and $T_2(v)\equiv \eins\sTP T(v)$, where
\begin{equation}
	T(u) \equiv \begin{pmatrix}
                    T^1_{~1}(u) & T^1_{~2}(u) \\
                    T^2_{~1}(u) & T^2_{~2}(u)
                \end{pmatrix}_{\!\! BF}
	     \equiv \begin{pmatrix}
                    A(u) & B(u) \\
                    C(u) & D(u)
                \end{pmatrix}_{\!\! BF}\ .
\end{equation}
The PBC super quantum determinant (SQD) is defined as
\begin{eqnarray}
	\delta\gklammer{T(u)} &\equiv& \sTr{12}{P_{12}^\sss{-}\ T_1(u)\ T_2(u+2\eta)} \\
	                      &=&      \frac{1}{2}\{ C(u)B(u+2\eta)-A(u)D(u+2\eta)\notag\\
	                      &~&      -B(u)C(u+2\eta)-D(u)A(u+2\eta)\}\ .
\end{eqnarray}
At $v=u+2\eta$ and after dividing by $a(2\eta)$ the graded YBA yields the
commutation relations 
\begin{eqnarray}
	C(u)B(u+2\eta)-A(u)D(u+2\eta) = C(u+2\eta)B(u)-D(u+2\eta)A(u)\label{eqAP:22}\\ 
	D(u)A(u+2\eta)+B(u)C(u+2\eta) = D(u+2\eta)A(u)-C(u+2\eta)B(u)\label{eqAP:23}\\ 
	B(u)C(u+2\eta)+D(u)A(u+2\eta) = B(u+2\eta)C(u)+A(u+2\eta)D(u)\label{eqAP:33}   
\end{eqnarray}
These relations can be used to simplify the super quantum determinant to
\begin{equation}
	\delta\gklammer{T(u)} = -[A(u)D(u+2\eta)-C(u)B(u+2\eta)]\ .
\end{equation}
It remains to show that the super quantum determinant is a central element of the graded YBA, i.e. that it supercommutes with all the other elements $A(v)$, $B(v)$, $C(v)$ and $D(v)$ for arbitrary $v$. Consider the expression
\begin{equation}
	R_{12}(u-v) R_{13}(u-w) R_{23}(v-w) T_1(u) T_2(v) T_3(w)\ .
	\label{eqAP:Proof1}
\end{equation}
Employing the graded YBE once it is obvious that
\begin{eqnarray}
	\text{(\ref{eqAP:Proof1})}   &=& R_{23}(v-w) R_{13}(u-w)\ \eckklammer{ R_{12}(u-v) T_1(u) T_2(v) }\ T_3(w)\\
	\boxed{v\rightarrow u+2\eta} &\Rightarrow& -2 a(2\eta) R_{23}(u-w+2\eta) R_{13}(u-w) P_{12}^\sss{-} T_1(u) T_2(u+2\eta) T_3(w)\ .
    \label{eqAP:ProofLHS}
\end{eqnarray}
On the other hand, by applying the graded YBA relation twice it is found that
\begin{eqnarray}
	\text{(\ref{eqAP:Proof1})}   &=& T_3(w)\ \eckklammer{ R_{12}(u-v) T_1(u) T_2(v) }\ R_{13}(u-w) R_{23}(v-w)\\
	\boxed{v\rightarrow u+2\eta} &\Rightarrow& -2 a(2\eta) T_3(w) P_{12}^\sss{-} T_1(u) T_2(u+2\eta) R_{13}(u-w) R_{23}(u-w+2\eta)\ .
    \label{eqAP:ProofRHS}
\end{eqnarray}
Equating (\ref{eqAP:ProofLHS}) and (\ref{eqAP:ProofRHS}) and multiplying from both sides with $P_{12}^\sss{-}$ gives
\begin{equation}
	\begin{split}
		\{ P_{12}^\sss{-} R_{23}(u-w+2\eta) R_{13}(u-w) P_{12}^\sss{-} \} \{ P_{12}^\sss{-} T_1(u) T_2(u+2\eta) P_{12}^\sss{-} \} T_3(w) \\
		 = T_3(w) \{ P_{12}^\sss{-} T_1(u) T_2(u+2\eta)  P_{12}^\sss{-} \}\{ P_{12}^\sss{-} R_{13}(u-w) R_{23}(u-w+2\eta) P_{12}^\sss{-} \}
	\end{split}
\end{equation}
where additional $P_{12}^\sss{-}$ projectors have been inserted by virtue of the appropriate triangularity conditions. After a change of basis to the $P_{12}^\sss{-}$ eigenbasis via $A_{12}$ as defined in (\ref{eq:HigherFusedRMatrices}), it is easy to check that application of the supertrace $\sTr{12}{.}$ yields
\begin{eqnarray}
	               &\qquad \sigma_3^z\ \delta\gklammer{T(u)}\ T_3(w)               &= T_3(w)\ \sigma_3^z\ \delta\gklammer{T(u)} \\
	\Leftrightarrow&\qquad \eckklammer{ \sigma_3^z\ \delta\gklammer{T(u)}, T_3(w)} &= 0 \label{eqAP:QDetCommutation}\\
	\Leftrightarrow&\qquad \eckklammer{ \delta\gklammer{T(u)}, T^i_{~j}(w)}_\pm    &= 0 \ .
\end{eqnarray}

Similarly one may introduce the object
\begin{equation}
	\delta\gklammer{\hat{T}(u)} \equiv \sTr{12}{P_{12}^\sss{-}\ \hat{T}_2(u)\ \hat{T}_1(u+2\eta)}
\end{equation}
which obeys the exact same super commutation relations and by
(\ref{eq:MonodromyHat}) turns out to be proportional to the inverse of the
above SQD. In particular for the considered $N$-site small polaron model it is
found that 
\begin{subequations}
\begin{eqnarray}
	\delta(u)       &\equiv& \delta\gklammer{T(u)} = -\zeta^N(u+2\eta)\ \prod_{i=1}^N (-\sigma^z_{q_i})     \label{eqAP:SQD}\\
	\hat{\delta}(u) &\equiv& \delta\{\hat{T}(u)\}  = -\frac{1}{\zeta^N(u)}\ \prod_{i=1}^N (-\sigma^z_{q_i}) \label{eqAP:hatSQD}
\end{eqnarray} 
\end{subequations}
where $q_i$ labels the $i$-th quantum subspace (cf. section \ref{ss:SuperTransfermatrixTruncation}). Moreover, the commutation relation (\ref{eqAP:QDetCommutation}) extends to the fused quantities according to
\begin{equation}
	\eckklammer{ \sigma_{\ll n\gg}^z\ \delta\gklammer{T(u)}, T_{\ll 1\ldots n \gg}^{\phantom{z}}(w)} = 0
	 \label{eqAP:fusedQDetCommutation}
\end{equation}
with $\sigma_{\ll n\gg}^z$ being defined in equation (\ref{eq:fusedSigmaZ}).

In the open boundary case, the place of $\delta\gklammer{T(u)}$ is taken by
another object $\Delta(u)$ which will most appropriately be called the OBC
super quantum determinant. Generally, the SQD is what you get when you alter
the first fusion step such that, instead of creating a \herv{higher
  dimensional} transfer matrix by projection on a three dimensional auxiliary
space, you now create a \herv{lower} dimensional object by projecting onto the
complementary one dimensional space. In a sense, loosely speaking, you do a
reduction instead of a fusion and find that the open boundary SQD factors as
follows,
\begin{equation}
  \begin{aligned}
    \Delta(u) &\equiv \sTr{12}{P_{12} K^\sss{+}_2(u+2\eta)
      \bar{R}_{12}(-2u-6\eta) K^\sss{+}_1(u) \mathcal{T}^\sss{-}_1(u)
      R_{12}(2u+2\eta) \mathcal{T}^\sss{-}_2(u+2\eta) }\\ 
    &= \delta\gklammer{K^\sss{+}(u)} \cdot \delta\gklammer{T(u)}
    \cdot \delta\gklammer{K^\sss{-}(u)} \cdot
    \delta\{\hat{T}(u)\}\\ 
    &= \klammer{\frac{\zeta(u+2\eta)}{\zeta(u)}}^N
    \delta\gklammer{K^\sss{+}(u)} \cdot
    \delta\gklammer{K^\sss{-}(u)} 
  \end{aligned}
  \label{eqAP:OBCQuantumDeterminant}
\end{equation}
where $\mathcal{T}^\sss{-}(u)$ was defined in (\ref{eq:RA_Representation}) and 
\begin{subequations}
  \begin{eqnarray}
    &&\begin{aligned}
      \delta\gklammer{K^\sss{+}(u)} &\equiv \sTr{12}{P_{12}^\sss{-}\
        K^\sss{+}_2(u+2\eta)\ \bar{R}_{12}(-2u-3\cdot 2\eta)\
        K^\sss{+}_1(u)}\\ 
      &= g(-2u-6\eta) \cdot
      \det\gklammer{K^\sss{+}(u)} 
    \end{aligned}\\
    &&\begin{aligned}
      \delta\gklammer{K^\sss{-}(u)} &\equiv \sTr{12}{P_{12}^\sss{-}\
        K^\sss{-}_1(u)\ R_{21}(2u+2\eta)\ K^\sss{-}_2(u+2\eta)}\\ 
      &= g(2u+2\eta) \cdot
      \det\gklammer{K^\sss{-}(u+2\eta)} 
    \end{aligned}
  \end{eqnarray}
\end{subequations}
with the function $g(u)$ being introduced in the context of
(\ref{eq:RMatrixPropertiesUnitarity}). Since
$\alpha_\sss{\pm}\cdot\beta_\sss{\pm} = 0$, as mentioned in section
\ref{ch:ReflectionAlgebrasAndBoundaryMatrices}, 
the determinants $\det\gklammer{K^\sss{\pm}(u)}$ depend \emph{only} on the
diagonal boundary parameters $\psi_\pm$.  This is different from the open XXZ
chain, where two parameters for each boundary enter the expression for the
quantum determinant.
%

\section{Transformation matrices}
\label{app:TrafoMatrices}
This appendix presents a collection of matrix representations of the various
similarity transformations employed in this paper. It is convenient to define
the coefficients 
%
%
\begin{subequations}
\begin{equation}
	a_n \equiv \sqrt{\frac{2n}{n+1} }\ \klammer{ [n]_{q}|_{\eta = \eta_n} }^{-1/2}
	\quad \text{and} \quad
	b = \klammer{\frac{[2]_q |_{\eta=\eta_2} }{[3]_q |_{\eta=\eta_3}}}^{-1/2}
\end{equation}
where $[n]_q$ denotes the usual $q$-deformation of an integer $n\in\mathbb{N}$ defined by
%
%
\begin{equation}
	[n]_q \equiv \frac{q^n-q^{-n}}{q-q^{-1}}
	\quad \text{with} \quad
	q \equiv \text{e}^{2\mathrm{i}\eta}
\end{equation}
and to set
\begin{equation}
	A_{(1)} \equiv B_{\ll 1\gg} \equiv C_{\ll 1\gg} \equiv \begin{pmatrix} 1 & 0 \\ 0 & 1 \end{pmatrix}\ .
\end{equation}
\end{subequations}
\begin{subequations}
\begin{eqnarray}
	A_{(12)} &=& \left( \begin{array}{cccc}
	   1 & 0 & 0 & 0 \\
	   0 & \frac{1}{\sqrt{2}} & \frac{1}{\sqrt{2}} & 0 \\
	   0 & 0 & 0 & 1 \\
	   0 & \frac{1}{\sqrt{2}} & -\frac{1}{\sqrt{2}} & 0
	\end{array}	\right) \\[.5em]
    B_{\ll 12\gg} &=& \operatorname{diag}(a_2,1,a_2)\\[.5em]
	C_{\ll12\gg} &=& \operatorname{diag}(a_2,1,1)
\end{eqnarray}
\end{subequations}

\begin{subequations}
\begin{eqnarray}
	A_{(123)} &=& \left( \begin{array}{cccccccc}
	   1 & 0 & 0 & 0 & 0 & 0 & 0 & 0 \\
	   0 & \frac{1}{\sqrt{3}} & \frac{1}{\sqrt{3}} & 0 & \frac{1}{\sqrt{3}} & 0 & 0 & 0 \\
	   0 & 0 & 0 & \frac{1}{\sqrt{3}} & 0 & \frac{1}{\sqrt{3}} & \frac{1}{\sqrt{3}} & 0 \\
	   0 & 0 & 0 & 0 & 0 & 0 & 0 & 1 \\
	   0 & \frac{2 \sqrt{2}}{3} & -\frac{\sqrt{2}}{3} & 0 & -\frac{\sqrt{2}}{3} & 0 & 0 & 0 \\
	   0 & -\frac{\sqrt{2}}{3} & \frac{2 \sqrt{2}}{3} & 0 & -\frac{\sqrt{2}}{3} & 0 & 0 & 0 \\
	   0 & 0 & 0 & \frac{2 \sqrt{2}}{3} & 0 & -\frac{\sqrt{2}}{3} & -\frac{\sqrt{2}}{3} & 0 \\
	   0 & 0 & 0 & -\frac{\sqrt{2}}{3} & 0 & \frac{2 \sqrt{2}}{3} & -\frac{\sqrt{2}}{3} & 0
	\end{array}	\right)\\[.5em]
	B_{\ll 123\gg} &=& \operatorname{diag}(a_3,1,1,a_3)\\[.5em]
	C_{\ll 123\gg} &=& \operatorname{diag}(a_3,1,1,1)
\end{eqnarray}
\end{subequations}

\begin{subequations}
\begin{eqnarray}
	A_{(1234)} &=& \frac{1}{\sqrt{2}} \left( \begin{array}{cccccccccccccccc}
	 \sqrt{2} & 0 & 0 & 0 & 0 & 0 & 0 & 0 & 0 & 0 & 0 & 0 & 0 & 0 & 0 & 0 \\
	 0 & \frac{1}{\sqrt{2}} & \frac{1}{\sqrt{2}} & 0 & \frac{1}{\sqrt{2}} & 0 & 0 & 0 & \frac{1}{\sqrt{2}} & 0 & 0 & 0 & 0 & 0 & 0 & 0 \\
	 0 & 0 & 0 & \frac{1}{\sqrt{3}} & 0 & \frac{1}{\sqrt{3}} & \frac{1}{\sqrt{3}} & 0 & 0 & \frac{1}{\sqrt{3}} & \frac{1}{\sqrt{3}} & 0 & \frac{1}{\sqrt{3}} & 0 & 0 & 0 \\
	 0 & 0 & 0 & 0 & 0 & 0 & 0 & \frac{1}{\sqrt{2}} & 0 & 0 & 0 & \frac{1}{\sqrt{2}} & 0 & \frac{1}{\sqrt{2}} & \frac{1}{\sqrt{2}} & 0 \\
	 0 & 0 & 0 & 0 & 0 & 0 & 0 & 0 & 0 & 0 & 0 & 0 & 0 & 0 & 0 & \sqrt{2} \\
	 0 & \frac{3}{2} & -\frac{1}{2} & 0 & -\frac{1}{2} & 0 & 0 & 0 & -\frac{1}{2} & 0 & 0 & 0 & 0 & 0 & 0 & 0 \\
	 0 & -\frac{1}{2} & \frac{3}{2} & 0 & -\frac{1}{2} & 0 & 0 & 0 & -\frac{1}{2} & 0 & 0 & 0 & 0 & 0 & 0 & 0 \\
	 0 & 0 & 0 & \frac{5}{3} & 0 & -\frac{1}{3} & -\frac{1}{3} & 0 & 0 & -\frac{1}{3} & -\frac{1}{3} & 0 & -\frac{1}{3} & 0 & 0 & 0 \\
	 0 & -\frac{1}{2} & -\frac{1}{2} & 0 & \frac{3}{2} & 0 & 0 & 0 & -\frac{1}{2} & 0 & 0 & 0 & 0 & 0 & 0 & 0 \\
	 0 & 0 & 0 & -\frac{1}{3} & 0 & \frac{5}{3} & -\frac{1}{3} & 0 & 0 & -\frac{1}{3} & -\frac{1}{3} & 0 & -\frac{1}{3} & 0 & 0 & 0 \\
	 0 & 0 & 0 & -\frac{1}{3} & 0 & -\frac{1}{3} & \frac{5}{3} & 0 & 0 & -\frac{1}{3} & -\frac{1}{3} & 0 & -\frac{1}{3} & 0 & 0 & 0 \\
	 0 & 0 & 0 & 0 & 0 & 0 & 0 & \frac{3}{2} & 0 & 0 & 0 & -\frac{1}{2} & 0 & -\frac{1}{2} & -\frac{1}{2} & 0 \\
	 0 & 0 & 0 & -\frac{1}{3} & 0 & -\frac{1}{3} & -\frac{1}{3} & 0 & 0 & \frac{5}{3} & -\frac{1}{3} & 0 & -\frac{1}{3} & 0 & 0 & 0 \\
	 0 & 0 & 0 & -\frac{1}{3} & 0 & -\frac{1}{3} & -\frac{1}{3} & 0 & 0 & -\frac{1}{3} & \frac{5}{3} & 0 & -\frac{1}{3} & 0 & 0 & 0 \\
	 0 & 0 & 0 & 0 & 0 & 0 & 0 & -\frac{1}{2} & 0 & 0 & 0 & \frac{3}{2} & 0 & -\frac{1}{2} & -\frac{1}{2} & 0 \\
	 0 & 0 & 0 & 0 & 0 & 0 & 0 & -\frac{1}{2} & 0 & 0 & 0 & -\frac{1}{2} & 0 & \frac{3}{2} & -\frac{1}{2} & 0
	\end{array}	\right)\\[.5em]
	B_{\ll 1234\gg} &=& \operatorname{diag}(a_4,1,b,1,a_4)\\[.5em]
	C_{\ll 1234\gg} &=& \operatorname{diag}(a_4,1,b,1,1)
\end{eqnarray}
\end{subequations}

\newpage
%

\end{document}